\documentclass[12pt, draftclsnofoot, onecolumn]{IEEEtran}

\newcommand{\nop}[1]{} 
\newtheorem{theorem}{Theorem}[section]
\newtheorem{lemma}[theorem]{Lemma}

\newtheorem{definition}{Definition}

\usepackage{listings}
\usepackage{flushend}
\usepackage{subfigure} 
\usepackage{scrextend}
\usepackage{hyperref}
\usepackage{wrapfig,lipsum,booktabs}

\usepackage{cite}

\ifCLASSINFOpdf
   \usepackage[pdftex]{graphicx}
\else
   \usepackage[dvips]{graphicx}
\fi
\usepackage{amsmath}

\usepackage{amssymb}

\usepackage{stfloats}
\fnbelowfloat
\usepackage{tikz}
\usetikzlibrary{shapes,arrows}
\usepackage[T1]{fontenc}

\begin{document}
%
\title{Towards Optimal Distributed Node Scheduling in a Multihop Wireless Network through Local Voting}
%
%
%

\author{Dimitrios~J.~Vergados,~\IEEEmembership{Member,~IEEE,}
        Natalia~Amelina,~\IEEEmembership{Member,~IEEE,}
        Yuming~Jiang,~\IEEEmembership{Senior Member,~IEEE,}
        Katina~Kralevska,~\IEEEmembership{Member,~IEEE,}
        and~Oleg~Granichin,~\IEEEmembership{Senior Member,~IEEE}
\thanks{D. J. Vergados is with the School of Electrical and Computer
        Engineering, National Technical University of Athens, Zografou, GR-15780, Greece, djvergad@gmail.com. }%
\thanks{N. Amelina and O. Granichin are with the Faculty of Mathematics and Mechanics,
Saint Petersburg State University, 7-9, Universitetskaya Nab., St. Petersburg, 199034, Russia, \{n.amelina, o.granichin\}@spbu.ru.}
\thanks{Y. Jiang and K. Kralevska are with the Department of Information Security and Communication Technology, 
Norwegian University of Science and Technology (NTNU), Trondheim, N-7491, Norway, \{jiang, katinak\}@ntnu.no.}
}

\maketitle

\vspace{-1.0cm}
\begin{abstract}
In a multihop wireless network, it is crucial but challenging to schedule
transmissions in an efficient and fair manner. In this paper, a novel
distributed node scheduling algorithm, called \emph{Local Voting}, is proposed.
This algorithm tries to semi-equalize the load (defined as the ratio of the queue
length over the number of allocated slots) through slot reallocation based on
local information exchange. The algorithm stems from the finding that the
shortest delivery time or delay is obtained when the load  is semi-equalized
throughout the network. In addition, we prove that, with Local Voting, the
network system converges asymptotically towards the optimal scheduling.
Moreover, through extensive simulations, the performance of Local Voting is
further investigated in comparison with several representative scheduling
algorithms from the literature. Simulation results show that the proposed
algorithm achieves better performance than the other distributed algorithms in
terms of average delay, maximum delay, and fairness. Despite being distributed,
the performance of \emph{Local Voting} is also found to be very close to a
centralized algorithm that is deemed to have the optimal performance.
\end{abstract}
\vspace{-0.2cm}

\begin{IEEEkeywords}
	\vspace{-0.2cm}
Multihop wireless networks, Node scheduling algorithm, Wireless mesh networks, Load balancing.
\end{IEEEkeywords}

%
\IEEEpeerreviewmaketitle

\section{Introduction}


Multihop wireless networks are a paradigm in wireless connectivity which
has been used successfully in a variety of network settings, including
ad-hoc networks~\cite{kiess2007survey}, wireless sensor
networks~\cite{pottie1998wireless}, and wireless mesh
networks~\cite{akyildiz2005wireless}. In such networks, the wireless devices may
communicate with each other in a peer-to-peer fashion and form a network, where
intermediate wireless nodes may act as routers and forward traffic to other
nodes in the network~\cite{bettstetter2005connectivity}.

Due to their many practical advantages and their wide use, there have been a lot
of studies on the performance of multihop wireless networks. For example, the
connectivity of a multihop wireless network has been studied under various
channel models in~\cite{gupta1998critical, bettstetter2005connectivity}.
Furthermore, their capacity  has been studied analytically
in~\cite{gupta2000capacity,li2001capacity,grossglauser2002mobility,
	weber2010overview}. In addition, the stability properties of scheduling policies
for maximum throughput in multihop radio networks have been studied
in~\cite{tassiulas1992stability, lin2004joint}. Also, a centralized scheduling
algorithm that emphasizes on fairness has been proposed in~\cite{salem2005fair}.
In~\cite{Ning2012533}, the authors focused on the joint scheduling and routing
problem with load balancing in multi-radio, multi-channel and multi-hop wireless
mesh networks. They also designed a cross-layer algorithm by taking into account throughput increase with load balancing.
Algorithms for joint power control, scheduling, and routing have been introduced
in~\cite{li2007joint, cruz2003optimal}. In~\cite{hyytia2006load},
the load balancing problem in a dense wireless  multihop network is formulated where the authors presented a general framework for analyzing the traffic load resulting from a given set of paths and traffic demands.

Some more recent literature works include~\cite{sgora2015survey,
  gunasekaran2010efficient,
	li2012efficient, chiang2014decentralized, arivudainambi2014heuristic,
	liu2012topology, zeng2014collaboration, xu2011topology, lam2014broadcast}.
In~\cite{sgora2015survey}, the authors present the state of the art in Time Division Multiple Access (TDMA)
scheduling for wireless multihop network.
Reference~\cite{gunasekaran2010efficient}
proposes Genetic Algorithm for finding Collision Free Set (GACFS) which is a co-evolutionary genetic algorithm
that solves the Broadcast Scheduling Problem (BSP) in order to optimize the slot assignment algorithm in WiMAX mesh networks.
It is a centralized approach and does not take into consideration the traffic
requirements or the load in the network.
Another scheduling solution for wireless mesh networks based on a memetic algorithm that does not consider the traffic requirements is presented in~\cite{arivudainambi2014heuristic}. An improved memetic algorithm is applied for energy-efficient sensor scheduling in~\cite{Arivudainambi2016}.
Reference~\cite{chiang2014decentralized} investigates the mini-slot scheduling
problem in TDMA based wireless mesh networks, and it proposes a decentralized
algorithm for assigning mini-slots to nodes according to their traffic
requirements.
The authors in~\cite{li2012efficient} propose a scheduling scheme for multicast communications where a conflict-free graph is created dynamically
based on each transmission's destinations.
Reference~\cite{liu2012topology} presents a probabilistic
topology transparent model for multicast and broadcast transmissions in mobile
ad-hoc networks.
The novelty of the scheme is that instead of guaranteeing that at least one conflict-free time slot is assigned to each node, it only
tries to bring the probability of successful transmission above a threshold.
The authors have further presented performance improvement for broadcasting in~\cite{6777543}.
Another topology transparent scheduling algorithm is presented
in~\cite{xu2011topology}.
The algorithm is not traffic dependent, and the achieved throughput is lower than the optimal mainly due to the requirement for a guaranteed slot for each node.
Reference~\cite{zeng2014collaboration} proposes a distributed scheduling scheme
for wireless sensor networks (WSNs).
Finally, the {\it NP}-hardness of the minimum latency broadcast scheduling problem is proved in~\cite{lam2014broadcast} under the Signal-to-Interference-plus-Noise-Ratio (SINR) model.
Two distributed deterministic algorithms for global broadcasting based on the SINR model are presented in~\cite{7524609}.

Efficient traffic load balancing and channel access are essential to harness the dense and increasingly heterogeneous
deployment of next generation 5G wireless infrastructure \cite{6824752}. Channel access in 5G
networks faces inherent challenges associated with the
current cellular networks \cite{Panwar201664}, e.g. fairness, adaptive rate control, resource reservation, real-time
traffic support, scalability, throughput, and delay. For instance, being able to do frequency
and time slot allocation enables more adaptive and sophisticated multi-domain interference management techniques \cite{Li:2013:OMB:2509723,7752514}. In \cite{7752514}, TDMA is used to mitigate the co-tier interference from time domain perspective in ultra-dense
small cell networks. 
The modeling and the optimization of load balancing plays a crucial role in the resource allocation in the next generation cellular networks \cite{6812287}.

In this paper, we focus on the problem of \emph{node scheduling} in multihop wireless networks.
In the node
scheduling problem, each transmission opportunity is assigned to a set of nodes
in a such way which ensures that there will be no mutual interference among any
transmitting nodes. More specifically, under node scheduling, two nodes can be
assigned the same time slot (and transmit simultaneously) if they do not have
any common neighbors. We introduce the \emph{Local Voting} algorithm.
The idea behind the algorithm was originated by the observation that
the total delivery time in a network can be minimized, if the ratio of
the queue length over the number of allocated slots is semi-equalized throughout the
network. We call this ratio the \emph{load} of each node.
The proposed algorithm allows for neighboring nodes to exchange slots in a
manner that eventually semi-equalizes the load in the network. The number of slots
that are exchanged is determined by the relation between the load of each node
and its neighbors, under the limitation that certain slot exchanges are not
possible due to interference with other nodes.
The preliminary results 
were presented in~\cite{vergados2017local}. This paper presents new algorithm and an
analysis of its performance, as well as new simulation results.
The simulation results of the comparative study between \emph{Local Voting} and other representative algorithms from the literature show that \emph{Local Voting} achieves the shortest end-to-end delivery time and greatest fairness compared to other distributed algorithms for different network densities. We also show that its performance is very close to a centralized algorithm.
The presented algorithm is a modification of the Local Voting protocol with
non-vanishing to zero step-size
which was suggested in~\cite{7047923}.
It belongs to the more general class of stochastic approximation decentralized
algorithms which have been studied early in~\cite{Tsitsiklis, Huang} with
decreasing to zero step-size.
However, changing the traffic parameters leads to an unsteady setting of the optimization problem.
For similar cases the stochastic approximation with constant (or
non-vanishing to zero step-size) is useful~\cite{Borkar08, GranAmelinaTAC15}.

The paper is organized as follows: Section~\ref{network model} describes thoroughly the network model. Section~\ref{proposedalgorithm} presents the proposed \emph{Local Voting} algorithm where Section~\ref{consensus} presents an analysis of the performance of the algorithm
in terms of achieving consensus. The simulation results in Section~\ref{evaluation} compare the performance of the proposed algorithm with other algorithms from the literature. Finally, Section~\ref{conclusion} concludes the paper.

\section{Network Model and Load Balancing}
\label{network model}

Consider a network that can be represented by a graph ${\cal G}=(N,E)$. $N$ is
the set of all wireless nodes that communicate over a shared wireless channel,
i.e. $N = \{1,2,\ldots,n\}$.
$E$ is the set of directional but symmetric edges which exist between two
nodes if a broadcast from one node may cause interference
on the other node. 
We use the terms edges and links interchangeably.
Access on the channel is considered to follow a
paradigm of time division multiple access.
There is no spatial movement of the nodes.

\begin{figure*}[thpb]
	\centering
	\includegraphics[width=0.48\textwidth]{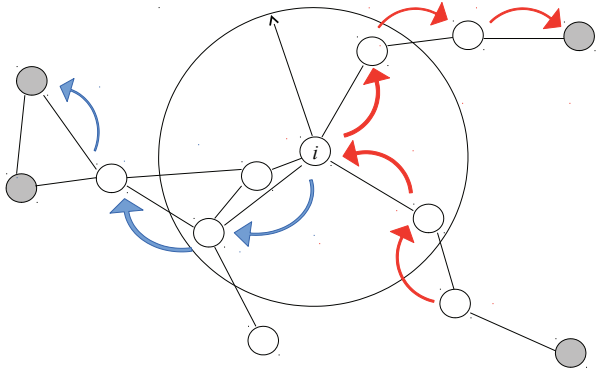}
	\vspace{-0.2cm}
	\caption{A multihop wireless network where the communication range and the interference range of node $i$ are denoted by the circle. The nodes with white background are two-hop neighbors of node $i$, and the nodes with gray background are outside the two-hop neighborhood of node $i$.}
	\label{refFigure}
	\vspace{-0.4cm}
\end{figure*}

The considered scheduling algorithm  is a \emph{node scheduling}
algorithm, i.e. each slot is allocated to a node, instead of a communication link.
We study a simple protocol interference model where two nodes are
one-hop neighbors as long as their distance is less than the communication range. The interference range is considered to be equal to the communication range, and both values are considered constant throughout the network.
A multihop network is presented in Fig.~\ref{refFigure} where the nodes within the circle of node $i$ are \emph{one-hop neighbors} of node $i$, and the \emph{one-hop neighborhood} of node $i$ is denoted by $N_{i}^{(1)}$. We also define $N_{i}^{(2)}$ as a \emph{two-hop neighborhood} of node $i$, i.e. the set of all the nodes that are neighbors to node $i$ or that have a common neighbor with node $i$. Since the inclusion $N_{i}^{(1)} \subset N_{i}^{(2)}$ holds, the nodes with white background in Fig.~\ref{refFigure} are \emph{two-hop neighbors} of node $i$.
The nodes presented with gray background are outside the two-hop neighborhood of node $i$.
Note that the nodes within the circle of node $i$ are also within the interference range of node $i$ because the interference range and the communication range are equal.
Two flows are depicted with red and blue arrows, respectively.
According to the protocol interference model, two nodes can be assigned the same transmission
slot, with no collision, as long as they do not have any common neighbors.
Otherwise, a collision would happen, resulting in data loss.
Node scheduling tries to guarantee that no such collision happens.

Each node contains a queue with packets to be transmitted, and the internal
scheduling on the queue is first-come-first-serve. The maximum length of each
queue is considered to be unbounded.
Each node also has a set of slots that have been assigned to it, and
neighboring nodes may exchange slots.

Time is divided into frames where each frame is denoted with $t$ and $t=0, 1, \dots$.
In addition, each frame $t$ is divided into time slots.
The
number of time slots in each frame is considered to be fixed and equal to $|S|$ where all time slots have the same duration.
The number of slots in a frame $|S|$ is considered to be large enough for
every node to be able to obtain at least one slot in each frame, if needed.
This value can be determined by the chromatic number of the graph,
where there is an edge between any two-hop neighbors in the original graph
$\cal G$. The Greedy Coloring Theorem provides an upper bound for this chromatic
number which is equal to $\max_{i \in N}|N^{(2)}_i| + 1$~\cite{chvatal1984perfectly}.
The duration of a time slot is sufficient to
transmit a single packet.

The transmission schedule of the network is defined as,
\begin{equation}
  \label{Djv_01}
  X_t^{i,s} = \left\{
  \begin{array}{rl}
    1, &\mbox{ if a slot } s \in S \mbox{ is assigned to a node } i \in N;\\
    0, &\mbox{ otherwise;}
  \end{array} \right.
\end{equation}
for $t \ge 0$, with $X^{i,s}_0 = 0$  by convention.

The transmission schedule is \emph{conflict-free}, if for any $t$,
\begin{equation}
\label{Djv_1}
 X_t^{i,s} X_t^{j,s} = 0, \forall s \in S, i \in N, j \in N_{i}^{(2)}, i \neq j.
\end{equation}
For each $i \in N$, let ${\tilde N}_{t}^i$ denote a set of such nodes $j$ that node~$i$ can exchange slots with node $j$ and the produced schedule remains conflict-free and $E_t$ denote the corresponding subset of edges.

The objective of this work is to design a load balancing node scheduling
strategy to schedule nodes' transmissions in such a way that the minimum maximal
(min-max) nodal delay is achieved.
We will study the following scheme of slot assignment and transmission of packets (see Fig.~\ref{fig1:pack_processing}).

\tikzstyle{decision} = [diamond, draw, fill=gray!20,
    text width=4.5em,
    text badly centered, node distance=3cm, inner sep=0pt]
\tikzstyle{block} = [rectangle, draw, fill=gray!20,
    text width=5em, text centered, rounded corners, minimum height=4em, node distance=3cm]
\tikzstyle{wblock} = [rectangle, draw, fill=gray!20,
    text width=8em, text centered, rounded corners, minimum height=4em, node distance=3cm]
\tikzstyle{line} = [draw, -latex']
\tikzstyle{cloud} = [text width=4.5em, text centered, draw, ellipse,fill=red!20, node distance=3cm,
    minimum height=2em]

\begin{figure*}[thpb]
    \centering
    \scalebox{0.85}{%
      \begin{tikzpicture}[node distance = 2cm, auto]
        \linespread{1}
        \tikzstyle{every node}=[font=\small]
        \node [cloud] (init) {For every node $i \in N$};
         \node [block, right of=init] (start) {Start with $q_t^i, p_{t-1}^i, u_{t}^i$};
         \node [block, right of=start
         ] (slots) {Release / Assign time slots };
      \node [block, right of=slots
         ] (process) {Transmit packets Get new packets};
         \node [block, right of=process] (compute) {Compute $u_{t+1}^i$};
        \node [cloud, right of=compute] (stop) {Start next frame $t+1$};
         \path [line] (init) --  (start);
         \path [line] (start) --  (slots);
         \path [line] (slots) --  (process);
          \path [line] (process) --  (compute);
          \path [line] (compute) --  (stop);
      \end{tikzpicture}
    }
\vspace{-0.4cm}
      \caption{Procedure of slot assignment and transmission of packets during frame $t$.}
    \label{fig1:pack_processing}
    \vspace{-0.4cm}
\end{figure*}
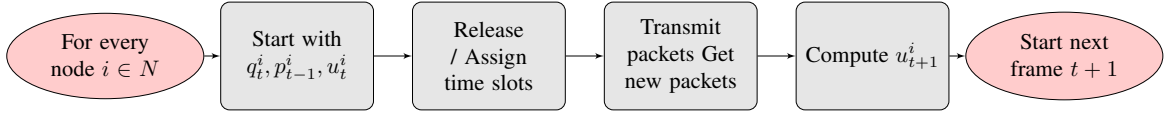

At the beginning of frame $t$, the state of each node $i$ in the network is described by three characteristics:
\begin{itemize}
  \item $q_{t}^{i} $ is the queue length, counted as the number of slots needed
    to transmit all packets at node~$i$ at frame $t$;
  \item $p_{t-1}^{i} $ is the number of slots assigned to node $i$ at the previous frame $t-1$,
    i.e. $p_{t-1}^{i} = \sum\limits_{s=1}^{|S|} X_{t-1}^{i,s} $;
  \item $u_{t}^{i}$ is the number of time slots which are assigned ($u_{t}^{i}>0 $) or released ($u_{t}^{i}<0 $) by node~$i$ at the beginning of frame $t$ ($u_{t}^{i}$ is calculated by the scheduling policy).
\end{itemize}
For each node $i$, the slot assignment starts with releasing time slots according to the scheduling policy when $u_{t}^{i}<0$, or otherwise with assigning slots to node $i$ from free time slots or through redistribution of time slots with its neighbors.
After that, the transmission of packets begins.
During frame $t$ new packets arrive.
At the end of frame $t$, the scheduling policy calculates $\{u_{t+1}^{i}\}_{i \in N}$ locally based on the available data.

So,
the dynamics of each node is described by
\begin{eqnarray}
\label{Nat_11}
\begin{aligned}
p_{t }^{i} & = p_{t-1}^{i} + n_t^i + u_{t}^{i},
\; i \in N,
\; t = 0,  1, \ldots, \\
q_{t + 1}^{i} & = \max\{0, q_{t}^{i} - p_{t}^{i}\} + z_{t}^{i},
\end{aligned}
\end{eqnarray}
where $n_t^i$ is the number of free slots that are allocated to node $i$ or
the number of slots that are released due to an empty queue, and $u_{t}^{i}$ is the number of time slots that node $i$
gains or loses at frame $t$ due to the adopted slot scheduling strategy.
These are slots that are exchanged between neighboring nodes, while $z_{t}^{i}$ is the number of slots needed to transmit new packets
received by node $i$ at frame $t$, either received as new packets from the
upper layers or from a neighboring node.
If $q_t^i=0$, then no slot is allocated
to the node $i$, i.e. we set $p_t^i=0$.


For reader's convenience, we provide Table \ref{notation} with the key notations used in this paper.
\begin{table}
  \renewcommand{\arraystretch}{0.7}                                             
  \caption{Table with notations}
  \begin{center} \label{notation}
	\begin{tabular}{ |c|l| }
  \hline
  ${\cal G}=(N,E)$ & Graph of a network topology \\
  $i$ & Node \\
	$N$ & Set of nodes in the network \\
$|N|$ & Number of nodes in the set $N$ \\
$E$ & Set of directional and symmetric edges between all two interfering nodes \\
$S$ & Set of slots in a frame \\
$|S|$ & Number of slots in a frame \\
$s$ & Time slot \\
$X_t^{i,s}$ & Transmission schedule for allocating slot $s$ to node $i$ at frame $t$ \\
$N_{i}^{(1)}$ & Set of one-hop neighbors of node $i$ \\
$N_{i}^{(2)}$ & Set of two-hop neighbors of node $i$ \\
$q_{t}^{i}$ & Queue length of node $i$ at frame $t$ \\
$p_{t}^{i}$ & Number of slots assigned to node $i$ at frame $t$ \\
$x_t^i$ & Load of node $i$ at frame $t$ \\
$z_{t}^{i}$ & Number of required slots to transmit new packets received by node $i$ at frame $t$ \\
$n_t^i$ & Number of free slots that are allocated to node $i$ or released due to an empty queue at frame $t$\\
$u_{t}^{i}$ & Number of slots that node $i$ gains or releases at frame $t$ \\
$\tilde N_t^i$ & Set of neighbors that can exchange slots with node $i$ at frame $t$ \\
$E_t$ & Set of edges between nodes that can exchange slots at frame $t$ \\
$A_t$ & Adjacency matrix corresponding to $E_t$\\
$a_t^{i,j}$ & Weight of edge $(j, i)\in E_t$ \\
${\cal G}_{A_t}$ & Graph defined by the adjacency matrix $A_t$ \\
$E_{\max}$ & Maximal set of communication links \\
$d^i(A)$ & Weighted in-degree of node $i$ (sum of $i$-th row of $A$) \\
$D(A)$ & Diagonal matrix of weighted in-degree of $A$ \\
${\cal L}(A)$ & Laplacian matrix of the graph ${\cal G}_A$ \\
$\lambda_1, \ldots, \lambda_n$ & Eigenvalues of the matrix ${\cal L}(A)$ \\
${\mathrm{E}}$ & Mathematical expectation \\
$\mathrm{E}_{{\cal F}_t}$ & Conditional mathematical expectation with respect to the $\sigma$-algebra ${{\cal F}_t}$ \\
$A_{av}$ & Adjacency matrix of the averaged system \\
$a_{av}^{i, j}$ & Mathematical expectation (average value) of $a_{t}^{i, j}$ \\
$\lambda_2(A_{av})$ & Second eigenvalue of the matrix $B_{av}$ ordered by absolute magnitude \\
$\left[ \cdot \right]$ & Round function \\
\hline
\end{tabular}
\end{center}
\end{table}

\subsection{Load Balancing} 

The ultimate objective of a scheduling algorithm in a multihop network is the packet flows to be delivered from the source to the destination in a
short time. This can be measured by the end-to-end delay per packet, the
end-to-end delivery time of a packet burst, the throughput of each flow, and
the fairness in distributing the resources among the competing flows.
In general, the problem of optimal scheduling in terms of approximating the
optimal throughput in a multihop wireless network is {\it NP}-hard as it is proven
in \cite{Jain:2003:IIM:938985.938993}.
A specific challenge of having such a scheduling algorithm is that it needs to
examine per flow information and use this information to schedule flows at
every node which we believe is difficult to implement.

For this reason, we do not optimize the end-to-end delay for the whole wireless
network, but instead we focus on optimizing the nodal (per-node) delay in each
transmitter.
The proposed \emph{Local Voting} algorithm may be considered as a compromise, where we
do node scheduling by using the slots without information about the individual flows.
Since multihop end-to-end delay is the sum of nodal delays on the end-to-end
path, we expect \emph{Local Voting} to deliver also good multihop end-to-end delay
performance.
To validate this, the evaluation in Section 4 has been focused on multihop
end-to-end delay, and the results indicate that \emph{Local Voting} does give good or
indeed better multihop end-to-end performance than various literature
algorithms.

In the following we show that the nodal delay may be optimized (min-max), if
the load of each node in the network is balanced.
The load of node $i$ at the beginning of frame $t$ is defined as zero
when $q^i_t=0$, and otherwise it is defined as the ratio of the queue length
$q^i_t$  over the number of allocated slots $p^i_t$ (note that slots are not assigned to nodes that have
nothing to transmit in an optimal scheduling strategy, so we have $q^i_t=0$ if $p^i_t=0$), i.e.
\begin{equation}
  x_i = \left\{
  \begin{array}{rl}
    \left[ \cfrac{q_i}{p_i} +0.5 \right], &\mbox{ if } q_i > 0, \\
    0, &\mbox{ if } q_i \mbox{ (and consequently } p_i) = 0.
  \end{array} \right. ,
\end{equation}
where
$\left[ \cdot \right]$ is the round function
(rounds a real number to the nearest integer).
Using this
definition we calculate the delay for each node $i$ (in time slots) as
$x_i \cdot |S|$.

\begin{definition}
Load balancing is the processes of equalizing the load between the nodes in the
network by exchanging slots among them.
\end{definition}

\begin{definition} We define a conflict-free schedule as ``nodally optimal''
or just ``optimal'', if the maximum delay per node in the network is
smaller or equal than the maximum per node delay for every other
schedule (min-max).
\end{definition}

\begin{lemma}
(Optimal schedules are maximal) An optimal schedule is a
(or has an equivalent) maximal schedule in the sense that
$\not \exists j \in N$ such that $p_j$ can be increased without reducing
$p_k$ in at least
one other node $k \in N$.
\end{lemma}

\begin{IEEEproof} Consider a schedule that is not maximal. That means there
exists $j \in N$ such that $p_j$ can be increased by one.
For the new schedule, the delay for all the other nodes is unchanged (since we did not
reduced slots for the other nodes). For node $j$, the new delay is $x'_i \cdot  |S| = \left[ \frac{q_i}{(p_i + 1)} +0.5 \right] \cdot  |S| \leq x_i \cdot  |S|$.
Thus, for every
non-maximal schedule, there exists a maximal schedule that has smaller
or equal maximum delay.
\end{IEEEproof}

\begin{lemma}\label{balanced}(Optimal schedules are balanced) Assume that node $k$ is the most
loaded node in the network, i.e $k = {\rm argmax}(x_i), i \in N$.
For all optimal schedules, it holds $x_k \leq x_j / (1-1/p_j)$ for the load of the most loaded node $k$ and the
load of every other node $j$ where $j\in \tilde N^k$.
\end{lemma}

\begin{IEEEproof} Assume that an optimal schedule exists where for the most
loaded node $k$, $x_k > x_j / (1-1/p_j)$ where $j \in \tilde N^k$.
Since $k$ is the most
loaded node, the maximal delay for such a schedule is $x_k \cdot  |S|$.
Since node $j \in \tilde N^k$, it follows that a slot of node $j$ can be
reassigned to node $k$. After reassigning, the new load for node $k$ is $[ q_k/(p_k +1) +0.5]$, and the corresponding delay for node $k$
is $\lceil q_k/(p_k +1) \rceil \cdot  |S| < x_k \cdot  |S|$. In addition, node
$j$ loses a slot so the new delay for node $j$ is $[ q_j/(p_j - 1) +0.5 ] \cdot  |S| = [ (q_j/p_j)/(1 - 1/p_j) +0.5 ] \cdot  |S|
= [(q_j/p_j)+0.5] /(1 - 1/p_j) \cdot  |S| = x_j/(1 - 1/p_j) \cdot  |S| <
x_k \cdot  |S|$. Thus, the new allocation has a maximal delay that is smaller than
or equal to the maximal delay of the other allocation, so the allocation is not optimal.
\end{IEEEproof}

Based on the above reasoning, we design a load balancing strategy
with two goals: 1) The produced schedule should be maximal, 2)
The load in the schedule should be balanced in the sense of Lemma \ref{balanced}.
For this reason, we define a slot exchange strategy that tries to equalize the load through load balancing, and in the next Section~\ref{consensus}
we prove that the \emph{Local Voting} algorithm converges to a such solution.

It should be noted that, in general, a schedule could be both maximal and
balanced, but still not optimal. This is because there could exist a
reallocation of the slots in the network that would produce a larger
spectral efficiency. Optimizing the schedule in this sense would require
finding a solution for the {\it NP}-complete broadcast scheduling problem. This is not easy, so for the
purposes of this paper, we do not examine ways of escaping local optima
and finding the global optimum. However, we can see from the simulation results that the performance of \emph{Local Voting} is still better than the performance of other distributed algorithms that we compare with, and also we see that
optimizing the maximal nodal delay also has a positive impact on the
end-to-end delay.

Among all possible options for load balancing, the min-max nodal delay is
achieved when all nonzero loads $q_t^i/p_t^i$ are semi-equal.
This comes as a result from the finding that the minimum expected nodal delay is achieved when the load in the network is equalized on nodes (Lemma 1 and Corollary 1 from \cite{7047923}).

\section{The Proposed Node Scheduling Algorithm: Local Voting}
\label{proposedalgorithm}

In the previous section
we have shown that an optimal schedule has three properties: it is efficient, it is
maximal, and it is balanced.
These are the properties which guide us in the design of the \emph{Local Voting}
algorithm.

In order to be efficient, there should be no slots allocated to nodes that have
an empty queue.
For this reason, before the beginning of each frame, nodes with an empty queue
release all time slots that they have reserved.

In order to be maximal, there should be no free time slot in the neighborhood
of any node, if that node has a positive queue, and assigning the slot to the node
would not cause a conflict with other nodes.
In order to meet this objective, after the first step, free slots are allocated to the nodes that do not
have an empty queue. Conflicts are resolved in a descending
order of the load.

Finally, the third objective is to be balanced, which can be formulated with
the following control goal:
 {\it to keep  the ratio $q_t^i / p_t^i$ semi-equal throughout the network
 (as much as possible) for the nodes $i$ where the queue is not empty $q_t^i >0$}.
In other words, the number of slots assigned to each node should correspond
to the amount of backlogged traffic.
A consequent implication is that, in order to achieve this optimal strategy,
we should be able to freely exchange slots among any two nodes in the network.
However, in reality, it is not always possible due to the potential
interference with other nodes in network. That is expressed through
Eq.~(\ref{Djv_1}).

In the following, we propose a novel algorithm that adopts the local voting
control strategy.
For the proposed \emph{Local Voting} algorithm, its semi-consensus properties with
respect to the local balancing
are
proved in
Section~\ref{consensus}.

\subsection{The Proposed Algorithm: Local Voting}

At the end of frame $t$, each node computes a scheduling policy.
The $u_{t+1}^i$ value is calculated as follows.

Each node uses the characteristics of its own state
$q_{t+1}^{i}$, $p_{t}^{i}$ and its neighbors' states
$q_{t+1}^{j}$, $p_{t}^{j}$ if $\tilde N_t^i \neq \emptyset$.

Let us for time frame $t$ and for each node $i, \, i\in N: \;q^{t+1}_i>0, $ define semi-inverse load $\tilde x_{t}^i$: $\tilde x_{t}^i = p^{t}_i / q^{t+1}_i$, and consider the following modification in the already known \emph{Local Voting (LV)}
protocol~\cite{7047923}:
\begin{equation}
  \label{Nat_7}
u_i^{t+1}=\left[ \gamma \sum_{j \in {\tilde N}_t^i} a_{t}^{i,j} (\tilde x_{t}^j - \tilde x_{t}^i) \right]
\end{equation}
where  $\gamma>0$ is a LV protocol step-size, and LV protocol matrix coefficients $a_{t}^{i,j}$: $$a_{t}^{i,j} = \frac {q_{t+1}^j}{1+\sum_{k \in {\tilde N}_t^i} \frac {q_{t+1}^k}{q_{t+1}^i}}.$$
Note, it is not so hard to see that
$$
u^i_{t+1}=\left[ \gamma \frac {\sum_{j \in {\tilde N}_t^i} q_{t+1}^i p_{t}^j - q_{t+1}^j p_{t}^i} { q_{t+1}^i + \sum_{j \in {\tilde N}_t^i} q_{t+1}^j} \right].
$$
For all other case we define $u_i^{t+1}=0$ and $\tilde x_{t}^i = p^{t}_i $. We set $a_{t}^{i, j} = 0$ for other pairs $i, j$ and denote the matrix of the
protocol as $A_t = [a_{t}^{i, j}]$. The
elements $a^{i,j}_t$
in  adjacency matrix $A_t$ are $a^{i,j}_t>0$ if node $i$ can exchange slots with node $j$ and the produced
schedule remains conflict-free; and $a^{i,j}_t=0$
 otherwise.

When $\gamma=1$, Eq.~(\ref{Nat_7}) has a form:
$$
  {u}_{t+1}^{i} = \left[ q^i_{t+1} \times \cfrac{p^i_{t}+\sum_{j\in {\tilde N}_t^i}{p^j_{t}}}{q^i_{t+1}+\sum_{j\in {\tilde N}_t^i}{q^j_{t+1}}}\right] - p^i_{t}.
$$

{\it Example}. Let's consider the network with $|S| = 50$, three nodes, all neighbors
with each other (single hop), with the following initial queue lengths
 of $q^1_0 = 400, \; q^2_0= 100, \; q^3_0= 310$, and $p^1_0 = 20,\; p^2_0=20,\;
 p^3_0=10$. The initial values for the loads are the following: $ x^1_0=20,\;
 x^2_0=5,\; x^3_0=31$.

The queue lengths at the end of frame $t=0$ will be $q^1_1=400-20=380,\;
 q^2_1=100-20=80,\; q^3_1=310-10=300$. Using Eq. (\ref{Nat_7}) we get
\begin{eqnarray}
\nonumber
u^1_1 = & 380 \cdot \left[ (20+20+10)/(380+80+300) \right] - 20 & = 5,\\
\nonumber
u^2_1 = & 80 \cdot \left[(20+20+10)/(380+80+300) \right] - 20 & = -15,\\
\nonumber
u^3_1 = & 300 \cdot \left[(20+20+10)/(380+80+300) \right] - 10 & = 10,
\end{eqnarray}
and we have three semi-equal loads
$$
x^1_1=380/25 = 15.2,\; x^2_1=80/5 = 16,\; x^3_1=300/20 \approx 15
$$


Eventually, node $i$ gains a slot in the following scenarios:
\begin{itemize}
  \item Its queue length is positive and there exists an available time slot
    that is not allocated to one-hop or two-hop neighbors of node $i$;
  \item Its queue length is positive and there exists a neighbor
  $j \in  {\tilde N}_t^i$ that has a value $u_{t}^{j}$ lower than zero.
\end{itemize}

It is important to note that the quantities in
protocol~(\ref{Nat_7}) are discrete-values, i.e. the state and other relevant
quantities may only take a countable set of values.
In that case, it makes sense to consider a quantised consensus
problem~\cite{kashyap2007quantized,kar2010distributed}.



\begin{figure*}[thpb]
  \begin{minipage}[b]{0.48\textwidth}
    \centering
    \scalebox{0.85}{%
      \begin{tikzpicture}[node distance = 2cm, auto]
        \linespread{1}
        \tikzstyle{every node}=[font=\small]
        \node [cloud] (init) {For every node $i \in N$};
        \node [decision, below of=init] (queue_empty) {Is queue empty? $q_t^i == 0$?};
        \node [decision, below of=queue_empty, node distance=9em] (found_free_slot) {Is there a free slot? };
        \node [decision, right of=queue_empty, node distance=15em] (allocated) {Are there allocated slots? $p_{t-1}^i > 0$?};
        \node [block, below of=found_free_slot] (balancing) {Load balancing};
        \node [block, below of=allocated, node distance=9em] (free_a_slot) {Release all slots};
        \node [block, right of=found_free_slot] (allocate) {Allocate $r$ free slots: $r \leq q_t^i$};
        \node [cloud, below of=free_a_slot] (stop) {End};

        \path [line] (init) --  (queue_empty);
        \path [line] (queue_empty.south) -- node{no} ++(0,-0.25) -| (found_free_slot.north);
        \path [line] (queue_empty) -- node {yes}(allocated);
        \path [line] (allocated) -- node {yes}(free_a_slot);
        \path [line] (allocated.east) -- node {no} ++(0.25,0) |- (stop.east);
        \path [line] (found_free_slot) -- node {yes}(allocate);
        \path [line] (allocate) -- (balancing);
        \path [line] (free_a_slot) -- (stop);
        \path [line] (found_free_slot) -- node {no}(balancing);
        \path [line] (balancing) -- (stop);

      \end{tikzpicture}
    }
    \caption{Requesting and releasing time slots function for the \emph{Local Voting} algorithm. }
    \label{fig:algo1}
  \end{minipage}
  \hfill
  \begin{minipage}[b]{0.48\textwidth}
    \centering
    \scalebox{0.85}{%
      \begin{tikzpicture}[node distance = 2cm, auto]
        \linespread{1}
        \tikzstyle{every node}=[font=\small]
        \node [cloud] (init) {Start};
        \node [block, below of=init, node distance=5em] (calcUi) {$u_t^i$ };
        \node [decision, below of=calcUi, node distance=9em] (ui_positive) {Is the control $u_t^i$ positive?};
        \node [cloud, below of=ui_positive, node distance=7em] (end) {End};
        \node [block, right of=calcUi,text width=9em, node distance=10em] (getslot) { Get $r$  slots: $r=\min\{u_t^i, u^i_t-u_t^j, p_{t-1}^j\}$ from node $j$, where $u_t^j = \min\limits_{m \in\tilde N^i_t } u_t^m $, $u_t^i:=u_t^i-r$, $u_t^j:=u_t^j+r$ };
        \node [decision, below of=getslot, node distance=9em] (available_slot) {Is there a node $j$ $\exists j \in {\tilde N}^i_t$ such that $u_t^j < 0 $? };

        \path [line] (init) -- (calcUi);
        \path [line] (calcUi) -- (ui_positive);
        \path [line] (ui_positive) -- node {no}(end);
        \path [line] (ui_positive) -- node {yes}(available_slot);
        \path [line] (available_slot) |- node {no}(end);
        \path [line] (available_slot) -- node {yes}(getslot);
        \path [line] (getslot) -- (calcUi);
      \end{tikzpicture}
    }
    \caption{Load balancing function for the \emph{Local Voting} algorithm.}
    \label{fig:algo2}
  \end{minipage}
\end{figure*}

The proposed \emph{Local Voting} algorithm consists of two  functions:
requesting and releasing free time slots, and load balancing.

For the first function (Fig.~\ref{fig:algo1}) nodes are examined sequentially
at the beginning of each frame.
If a node has an empty queue, then it releases all its time slots.
If a node has a positive backlog (i.e. its queue is not empty),
then it is given time slots. All time slots are examined
sequentially, and the first available time slots that are found, which are not
reserved by one-hop
or two-hop neighbors for transmission, are allocated to the node.
The message exchanges for requesting and releasing slots are considered equivalent to
message exchanges in the DRAND algorithm~\cite{rhee2006drand}.
If no available slot is found (all slots have been allocated to one-hop or two-hop neighbors of
the examined node), then no new slot is allocated to the node. On the contrary,
if the queue of the node is found to be empty and the node has allocated slots,
then all slots are released.

The \emph{load balancing} function (Fig.~\ref{fig:algo2}) is invoked in order
to achieve the objective of keeping the load balanced.
Every node $i\in N$ has a value $u_t^i$ (from the scheduling policy calculated at
the end of the previous frame) which determines how many
slots the node should ideally gain or lose by the load balancing function.
If a node has a
positive $u_t^i$ value, then it checks if any of its neighbors has a load lower that its own and may give a slot to it without causing a conflict.
Note that this is not always the case, because the requesting node may not be
able to obtain a slot if one of its other one-hop or two-hop neighbors has also
allocated the same slot.
The neighbor with the smallest $u_t^j$ value gives slots
to node $i$.
After the exchange $u_t^i$ is reduced by $r = \min\{u_t^i, u^i_t-u_t^j, p_{t-1}^j\}$,
and $u_t^j$ is increased by $r$.
This procedure is repeated until $u_t^i$ is  positive, or
until none of the neighbors of node $i$ can give any slots to node $i$ without
causing a conflict.
In this way, in general, slots are removed from nodes with lower load and are
offered to nodes with higher load, and eventually the load between
nodes will reach a common value, i.e. semi-consensus will be achieved.

\subsection{Consensus Properties of Local Voting}
\label{consensus}

\subsubsection{Notation}

For the considered network, $N_i^{(1)}$ and $ N_i^{(2)}$ do not change over time
since there is no spatial movement of the nodes.
However, the network changes over time due to the slot allocation which is dynamic.
Taking this into consideration, we describe the structure of the dynamic network
(network topology) using a sequence of directed graphs
${\cal G}_{A_t}=\{(N, E_{t})\} _{t\ge 0} $, where $E_{t} \subseteq E$.
In the considered case, $E_{t}$ defines a subset which consists of links between
the nodes that can exchange slots at time $t$.
Note that these
directed graphs ${\cal G}_{A_t}$ are not the same as the communication graph
${\cal G}$.
Instead, they define to which of the other nodes a node can offer a slot.
More specifically, if there is an edge from node $i$ to node $j$ in ${\cal G}_{A_t} $,
it means that node~$i$ has a slot to offer to node~$j$, and after the exchange
the produced schedule will still remain conflict-free with respect to Eq.~(\ref{Djv_1}).

$A_t = [a^{i, j}_t]$ is the corresponding adjacency matrix.
As defined earlier, ${\tilde N}_t^i = \left\{ j: a_t^{i, j} > 0 \right\}$
denotes the set of neighbors of node $i \in N$ at time $t$, i.e.
the set of neighbors that can exchange slots with node $i$.
Generally, $ {\tilde N}_t^i \neq \emptyset$~~if~~$\exists s \in S: X_t^{i,s} = 1$ and
$\forall k \in N_{i}^{(2)} \cup {i}, \; \; X_t^{i,s} X_t^{k,s} =0$.
Note that in
contrast to $N_{i}^{(1)}$ and $N_{i}^{(2)}$, the set $ {\tilde N}_t^i \subset N_{i}^{(1)}$
changes in time.
Let $E_{\max } = \{ (j, i): \sup_{t\ge 0} a_t^{i,j} > 0\}$ stand for the
maximal set of communication links (a set of edges that appear with non-zero
probability in ${\tilde N}_t^i$). For any matrix $A$ we define the weighted in-degree of node $i$ as a sum
of $i$-th row of the matrix $A$: $d^i(A)=\sum_{j=1}^n a^{i, j}$, and
$D(A) ={\rm diag}\{d^i(A)\}$ as the corresponding diagonal matrix. Let
${\cal L}(A) = D(A) - A$ denote the Laplacian matrix of the graph ${\cal G}_A$, and
$\lambda_1, \ldots, \lambda_n$ stand for the eigenvalues of the matrix
${\cal L}(A)$ ordered by increasing absolute magnitudes. The symbol $d_{\max}(A)$ accounts for a maximum in-degree of
the graph~${\cal G}_A$.

\vspace{2mm}
\subsubsection{Assumptions}

Let $({\Omega},{\cal F},{P})$ be the underlying probability space corresponding
to the sample space, the collection of all events, and the probability measure,
respectively, and $\{{\cal F}_t\}$ be a sequence of $\sigma$-algebras which are
generated by $\{q_k^i,p_k^i\}_{i=1, \ldots, n, k=1, \ldots, t}$. The symbol ${\mathrm{E}}$
accounts for the mathematical expectation, $\mathrm{E}_{{\cal F}_t}$ is a
conditional mathematical expectation with respect to the $\sigma$-algebra
${{\cal F}_t}$, and the following assumptions are satisfied:

\vspace{2mm}

{\bf A1}.
{\bf a)}
For all $i \in N, j \in N^{i}_{\max}$ an appearance of ``variable'' edges $(j, i)$
in the graph ${\cal G}_{A_t}$ is an independent random event. $N^{i}_{\max}$ is defined by the topology $E_{\max}$.

Denote by $a_{av}^{i, j}$ the average value of $a_{t}^{i, j}$. Let $A_{av}$
stand for the adjacency matrix of  averaged values $a_{av}^{i, j}$.

{\bf b)}
For all $i \in N,\; t=0, 1, \ldots$, the number of slots $z_{t}^{i}$ required to transmit new packets received by node $i$ at frame $t$
in Eq. (\ref{Nat_11}) are
random variables do not depend on~${\cal F}_{t}$.

Note that new packets refer to new incoming packets from new connections and new packets arrived from neighbors.


{\bf c)}
For all $i \in N,\; j \in {\tilde N}_t^i$ and  $b_{t}^{i, j}= \frac {q_{t+1}^j}{{q_{t+1}^i}+\sum_{k \in {\tilde N}_t^i} {q_{t+1}^k}}$ there exist conditional average values $b_{av}^{i, j} = \mathrm{E}_{{\cal F}_{t-1}}
(b_{t}^{i, j})$, which do not depend on $t$.
Note that $b_{t}^{i, j}=a_{t}^{i,j} /{q_{t+1}^i}$ and $B_{t}= A_{t} Q_{t+1}^{-1}$ where  $Q_{t+1}={\rm diag}\{\max\{1, q^i_{t+1}\}\}$.

There exists a positive definite matrix $Q_{av}>0$ such that
$A_{av}=B_{av}Q_{av}$, and ${\rm E}\|Q_{t+1}^{-1}-Q_{av}^{-1}\|^2 \leq \sigma_q^2$.

{\bf d)} For matrices $B_{t}=[b_{t}^{i, j}]$ and $B_{av}=[b_{av}^{i, j}]$  there exists a matrix $R$ such that
$$ \mathrm{E} ({\cal L}(B_{av})-{\cal L}({B}_t))^{\rm T}({\cal L}(B_{av})-{\cal L}({B}_t)) \leq R,$$
and its maximum on the absolute magnitude eigenvalue:
 $\lambda_{\max}(R)<\infty$.
 
{\bf e)}
For all $i \in N,\,t=0,1,\ldots$, the errors of rounding in LV protocol~(\ref{Nat_7})
\begin{equation}\label{eq_w}
w_{t}^i =  \gamma \sum_{j\in \tilde N_{t}^{i}} a_{t}^{i, j} (\tilde x_{t}^{j} - \tilde x_{t}^{i}) - [ \gamma \sum_{j\in \tilde N_{t}^{i}} a_{t}^{i, j} (\tilde x_{t}^{j} - \tilde x_{t}^{i}) ]
\end{equation}
 are centered, independent, and they have a bounded variance $\mathrm{E} (w_{t}^i)^2 = \sigma_w^2$ and independent of~${\cal F}_{t}$.

{\bf f)}
For all $i \in N,\,t=0,1,\ldots$, the variables $e_{t+1}^i$ are random, independent and identically distributed with mean values $\bar e^i$ and variance $\sigma_e^2$, and they do not depend on~${\cal F}_{t}$.

All variables $z_t^i, e_{t+1}^i, w_{t}^i$ are mutually independent.

\vspace{2mm}

We assume that the following assumption for the average matrix of the network
topology is satisfied:

{\bf A2:} Graph $ {\cal G}_{A_{av}}$ has a spanning tree, and for any edge
$(j, i) \in {E}_{\max}$ it holds $a_{av}^{i, j} > 0$.

\vspace{2mm}
\subsubsection{Mean Square $\epsilon$-consensus}
Consider the state vectors ${\tilde {\mathbf{x}}}_t \in {\mathbb R}^{n}, \; t=0,1, \ldots,$ which consist of the elements $\tilde x_t^1, \tilde x_t^2,  \ldots, \tilde x_{t}^{n}$. Note that if state values $\tilde x_t^i,\; i \in N,$ are semi-equal then the inverse values $q_{t+1}^i/p_t^i,\; i \in N$ for $q_{t+1}^i>0, p_t^i>0$ are semi-equal.

The following theorem gives the conditions when  the sequence
$\{{\mathbf{x}}_t \}$ converges asymptotically  in the mean squared sense to some bounded set around   a trajectory 
$\bar{\mathbf{x}}_{t} $ of the corresponding averaged model
\begin{equation}\label{aver_syst}
    {\bar {\mathbf{x}}}_{t+1} = {\bar {\mathbf{x}}}_{t} - \gamma {\cal L}({B}_{av}){\bar {\mathbf{x}}}_t+Q_{av}^{-1}\bar {\bf e}_{t+1},\;{\bar {\mathbf{x}}}_0 = 0 (={ {\mathbf{x}}}_0).
\end{equation}

If $\bar {\bf e}_t \equiv 0$ then  $\bar{\mathbf{x}}_{t} \to \bar{\mathbf{x}}_{\star} $ as $t \to \infty$, and  $\bar{\mathbf{x}}_{\star} $ is a left eigenvector of the matrix ${A}_{av}$
corresponding to its zero eigenvalue. 
Note that if $A_{av}$ is a symmetric matrix, then $\bar{\mathbf{x}}_{\star} $
is equal to $x_{\star} \mathbf{1}_{n}$ where $\mathbf{1}_{n}$ is $ n$-vector of
ones, i.e. we will get the asymptotical consensus for the state vectors
$\{\bar{\mathbf{x}}_t \}$.

{\bf Theorem 1.} {\bf If} Assumptions {\bf A1--A2} are satisfied and
\begin{equation}\label{cond_gamma1}
     0 < \gamma <  \frac{1}{d_{\max}(B_{av})},
\end{equation}
{\bf then}
\begin{equation}\label{cond_gamma2}
   \rho = (1-\lambda_2(B_{av}))^2 < 1
\end{equation}
and the trajectory $\{\bar{\mathbf{x}}_t \}$ of the system~(\ref{aver_syst}) converges
to the vector $\bar{\mathbf{x}}_{\star} $  which is a left eigenvector of the matrix
${A}_{\max}$ corresponding to its zero eigenvalues, and the following
inequality holds:
\begin{equation}
 \label{Nat_T1}
{\mathrm{E}} \|{\tilde {\mathbf{x}}}_{t+1} - {\bar{\mathbf{x}}}_{t+1} \|^2 \leq
2 (\frac{\Delta}{1-\rho}+  \rho^t {\mathrm{E}} \|{{\mathbf{p}}}_{0} \|^2 +\sigma_q^2\| Q_{av} {\bar{\mathbf{x}}}_{t+1}\|^2),
 \end{equation}
where
$$
\Delta = n(\lambda_{\max}(R) |S| + \sigma_e^2 + \sigma_w^2).
$$

{\bf If} $t \to \infty$,
 {\bf then} the asymptotic mean square $\varepsilon$-consensus
 is achieved with
$$
\varepsilon \leq 2 \frac{\Delta}{1-\rho} + 2\sigma_q^2\| Q_{av} {\bar{\mathbf{x}}}_{t+1}\|^2.
$$

Proof is in the Appendix.


Theorem 1 shows that our protocol~(\ref{Nat_7}) provides an approximate consensus,
i.e. gives an almost optimal behavior of the system.

\section{Evaluation}
\label{evaluation}

We have performed a set of simulations in order to evaluate the performance of different scheduling algorithms. These simulations are carried out by using a
custom--built, event-driven simulation tool developed in Java. The simulation
setup is summarized in Table~\ref{parameters}. \nop{ The network consists of $100$
nodes, randomly placed in the area of $100\times100$ units using a two-dimensional
uniform distribution. The topologies that are not entirely connected are discarded. Both the transmission range and the interference range are equal to $10$ units. With this setup, each node has on average $18.8$ one-hop neighbors
and $15.2$ two-hop neighbors, and the diameter of the corresponding
communication graph is on average $11.2$. Each TDMA frame consists of $10$ time
slots (for the algorithm with fixed frame length) where the duration of each time slot is $1$ unit of time.}


\begin{wraptable}[10]{r}{10cm}  
	\renewcommand{\arraystretch}{0.7}
	\label{parameters}
	\centering
	\caption{Simulation Setup}
	\begin{tabular}{l  l}
		\hline
		\bfseries Parameter & \bfseries Value\\
		\hline
		Number of Nodes & 100\\
		Transmission/Interference range & 10 units \\
		Topology size & 100 x 100 units \\
		Frame length & 10 time units \\
		Number of concurrent connections & 1 - 30 \\
		Number of packets per connection & 100 \\
		Packet generation interval & Every 5 slots \\
		Number of iterations & 500 \\
		\hline
	\end{tabular}
\end{wraptable}

Although several routing algorithms for load balancing in multihop networks
exist, e.g.~\cite{vergados2012fair}, in this paper we focus on the interaction
of scheduling and load balancing algorithms. The routing in the network
is considered to follow a simple shortest path routing algorithm.

\subsection{The Simulation Tool}

The source code that was developed for evaluating different scheduling
algorithms has been made open source and is available.\footnote{\url{https://github.com/djvergad/local_voting_src}} 
The scripts for running
the simulations and producing the results have also been made available.%
\footnote{\url{https://github.com/djvergad/local_voting}}

The simulation tool focuses on the evaluation of the scheduling algorithms.
There are two types of scenarios that were evaluated. In the first class of
scenarios, a variable number of connections is considered, each connection 
starts with a fixed number of packets. This represents the response to a 
sudden burst of 
traffic. 
Different load in the network is calibrated by changing the number
of connections. 
The simulation is executed until all packets have reached 
their destinations. 
In the second class of scenarios, connections are added 
constantly, following a Poisson process. The load is calibrated by changing 
the connection arrival rate. 
This scenario is executed for a fixed time
duration.

The measured metrics for each connection are:
\begin{itemize}
  \item the delivery time, which is the time needed for all packets of a
        connection to reach their final destination;
  \item The delay, which is time from the moment each packet is generated 
        until it has been received by its final destination;
  \item The throughput, which is the number of packets in the connection, 
        divided over the time difference (in slots) between the start and
        the completion of the connection.
\end{itemize}

For each simulation we used the per connection metrics in order to take 
the average value between the connections per simulation, the maximum and 
minimum values for each connection, and the fairness, which was calculated
using Jain's fairness index~\cite{Jain:1984:QMF}.

The simulation software is organized into four packages: the network package
contains the implementation of the network elements and algorithms, the
simulator package which contains the objects for implementing the 
discrete--event simulator, the application package which implements the 
network connections and the statistics gathering functionality, and the 
stability package which contains the different scenarios to be executed.

Some of the network functions that were implemented in the simulation tool 
include the following: a Connection object represents the application layer.
For the purposes of this simulation, each connection has a random source and 
destination. It is initialized with a number of packets that are transmitted.
For the first scenario (traffic bursts), each connection has 100 packets. For 
the steady state scenario, the number of packets are calculated based on
an exponential distribution. The Node object represents each wireless station
in the network. It contains an infinite FIFO queue that is common for all 
outgoing transmissions. It also has a routing table that is created using 
a shortest path algorithm. It contains a set of slot reservations, as well
as X-Y coordinates. A Reservation object represents the slot reservation. It 
contains fields for the transmitting node, as well as the nodes that are 
blocked due to this reservation (all nodes in the two-hop neighborhood, except
for the link-scheduling case). The Network object implements network functions,
such as routing. The Scenario object contains the scenario to be executed,
and defines the scheduler type, the transmission range, the number of time slots
in each frame, the number of nodes in the network, and the size of the topology.
Each Scheduler also has a different class which inherits from the 
TDMAScheduler class. The wireless channel is lossless (unless otherwise 
specified). Two nodes are one-hop neighbors if their distance is smaller than
the transmission/interference range. All scheduling algorithms are 
conflict-free using the protocol interference model where two nodes are not
scheduled to transmit as long as they are two-hop neighbors. We also consider a scenario with a link-scheduling algorithm where two transmissions are allowed to be concurrent, if each
receiver receives at most one packet at a time. 

\subsection{Implemented Algorithms}
\label{sec:algos}

In this subsection we briefly describe the operation of some algorithms for node
scheduling from the literature. We have implemented these algorithms in our 
simulation platform, and compared their performance with the performance of \emph{Local voting} algorithm.

A typical example of a distributed, traffic independent, topology dependent node
scheduling algorithm is DRAND~\cite{rhee2006drand}. DRAND defines a
communication protocol for obtaining a conflict-free schedule, using information
from the two-hop neighborhood. The protocol assigns a single time slot to each
node. The frame length is constant throughout the network, and it is determined by
the maximum density of the nodes.

Another example of a distributed, traffic independent, topology dependent node
scheduling algorithm is Lyui's algorithm~\cite{Lyui,hammond2004properties}. The
algorithm first assigns a color to each node, using existing graph coloring
techniques, with the limitation that two nodes are not assigned the same color
if they are in the same two-hop neighborhood. Depending on the color that is
assigned to a node, it is a candidate to transmit in any time slot for which
$t\bmod p(c_k) = c_k \bmod p(c_k)$, where $t$ is the time slot, $c_k$ is the
color assigned to node $k$, and $p(c_k)$ is the smallest power of $2$ greater
than or equal to $c_k$. 
Among these candidate nodes, in each two-hop neighborhood, the node with the largest color transmits.
Therefore, in Lyui's algorithm, the nodes have more than one transmission
opportunity in each frame, and there is no common frame length for the entire
network. This makes slot assignment easier than in DRAND where the frame length
must be known in advance. Lyui's algorithm also has better performance since
the nodes can transmit more frequently, and the performance in sparse areas is
not affected by larger node density elsewhere.

The \emph{Load-Based Transmission Scheduling}
(LoBaTS)~\cite{wolf2006distributed} protocol is an example of a distributed,
traffic dependent, topology dependent node scheduling scheme. It schedules the
transmissions using Lyui's algorithm, but now instead of each node having a
single color, additional colors can be assigned to nodes that experience high
load. Each node maintains an estimate of the utilization of every node in its
two-hop neighborhood. If the queue length exceeds a threshold, then the node
tries to find an additional color that: a)~is not assigned to any other node in
the two-hop neighborhood, and b)~does not cause the utilization of any other
node in the neighborhood to exceed one. If such a color is found, then the node
informs its neighbors about the new assignment, and it uses Lyui's algorithm to
calculate the new transmission schedule.

A centralized, traffic dependent, topology dependent node scheduling algorithm
was proposed in~\cite{dimakis2006sufficient}, called \emph{Longest Queue First}
(LQF) scheduling. According to this scheduling algorithm, nodes that have a
packet to transmit are ordered according to their queue length in a descending
order. The node with the longest queue is assigned to transmit in the current
time slot. The remaining nodes are examined one by one, and any node that can
transmit in the same time slot without causing a conflict is also assigned to
transmit. The LQF policy is a simple heuristic for slot assignment, but it is
not really practical, since it is centralized and the scheduler requires
information about the queue lengths of all nodes in the network. Nevertheless, due
to its simplicity and good performance, this algorithm has been often used for
obtaining theoretical results and as a benchmark for comparing the performance
of scheduling schemes. This algorithm is also known as the \emph{Greedy Maximal
	Scheduling} algorithm, and its performance in terms of capacity has been
analyzed in \cite{joo2009understanding}.

For the final scenario we used a link-scheduling variant of the LQF algorithm.
In this version of the algorithm, again the nodes are examined in decreasing
queue size. This time, however, whether the packet will conflict with other
transmissions depends on the destination of the packet (since we have link 
scheduling). For this reason, we examine the packets from the start of the
queue until we find the first packet that has a destination that doesn't cause
a convict with the already scheduled transmissions in this slot. This packet is
added to the slot, and the algorithm continues with the next node.

\subsection{Delivery Time Scenario}
\label{sec:deliv}

In this experiment we investigate the delivery time of fixed sized messages, all 
initialized at the same time.  
The scenario has been repeated $500$ times for each number of connections and
for each of the algorithms. The total number of
experiments is $500 \times 30 \text{ connections} \times 5 \text{ protocols} =
75000 \text{ experiments}$.

At the beginning of each simulation a varying number from 1 to 30 concurrent
connections is generated with random sources and destination nodes. Each
connection generates 1 packet every 5 time units until a total of 100 packets
per connection is generated.

\begin{figure*}[thpb]
	\centering
	\subfigure[The average end to end delay]
	{\includegraphics[width=0.48\textwidth]{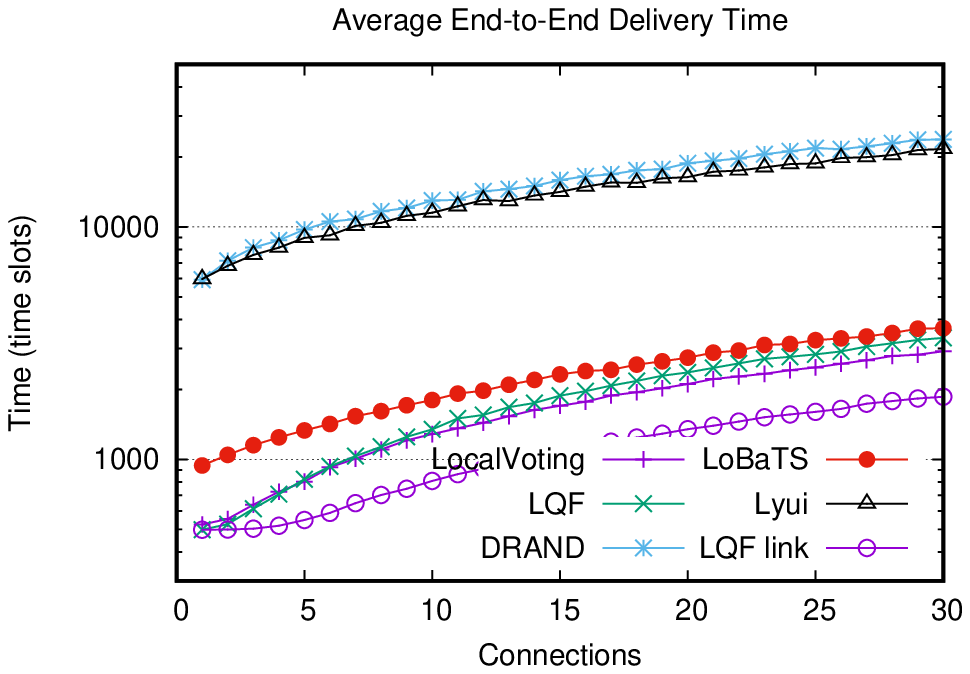}} 
	\hfill
	\subfigure[The fairness in end-to-end delay]
	{\includegraphics[width=0.48\textwidth]{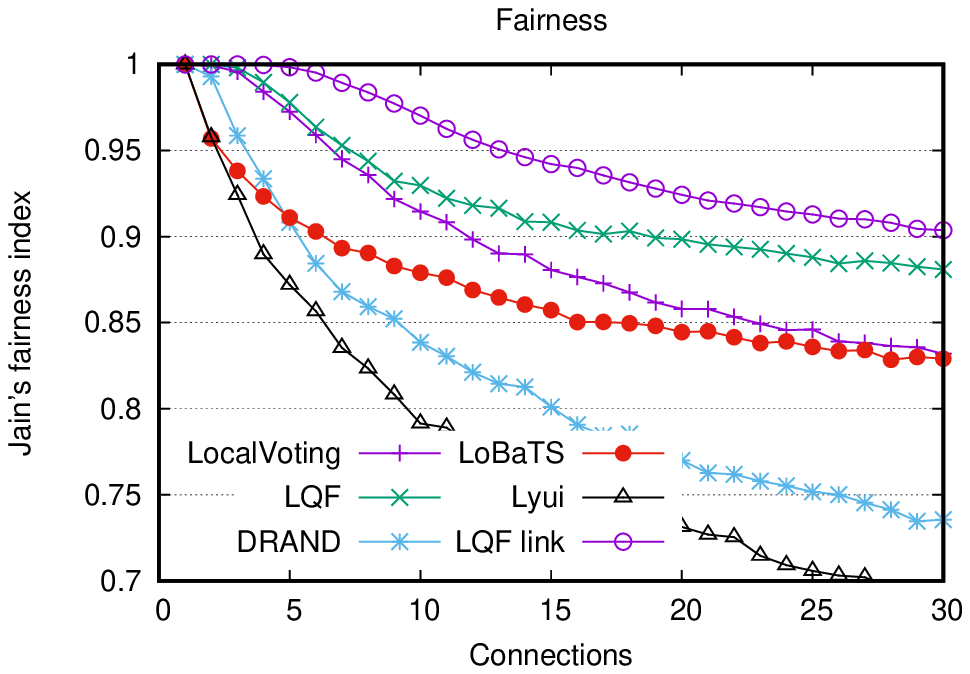}} 
	\subfigure[The maximum end-to-end delay]
	{\includegraphics[width=0.48\textwidth]{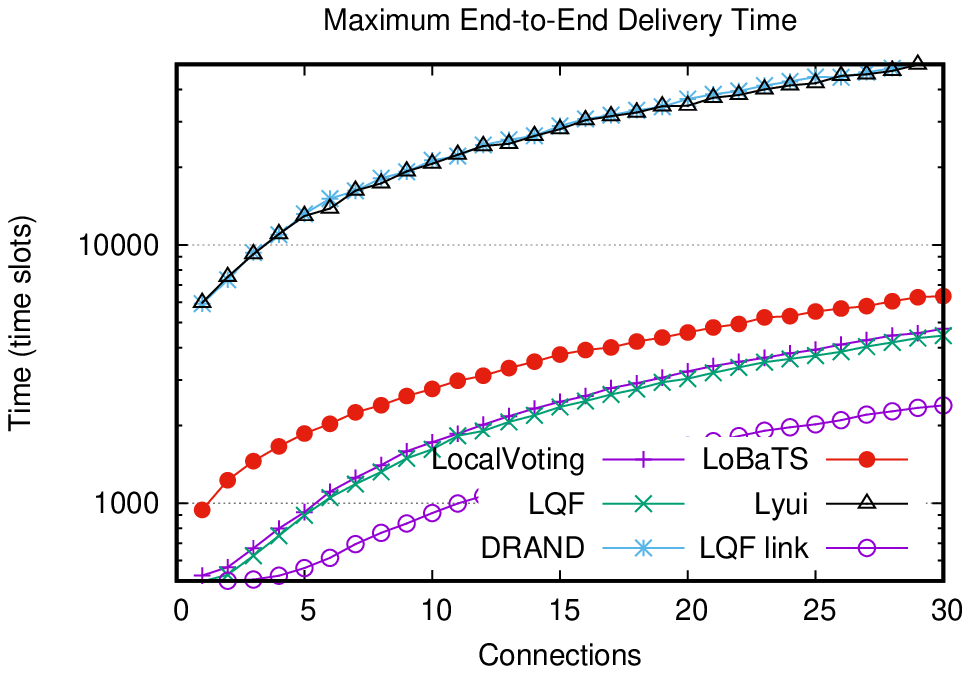}} 
	\hfill
	\subfigure[The minimum end-to-end delay]
	{\includegraphics[width=0.48\textwidth]{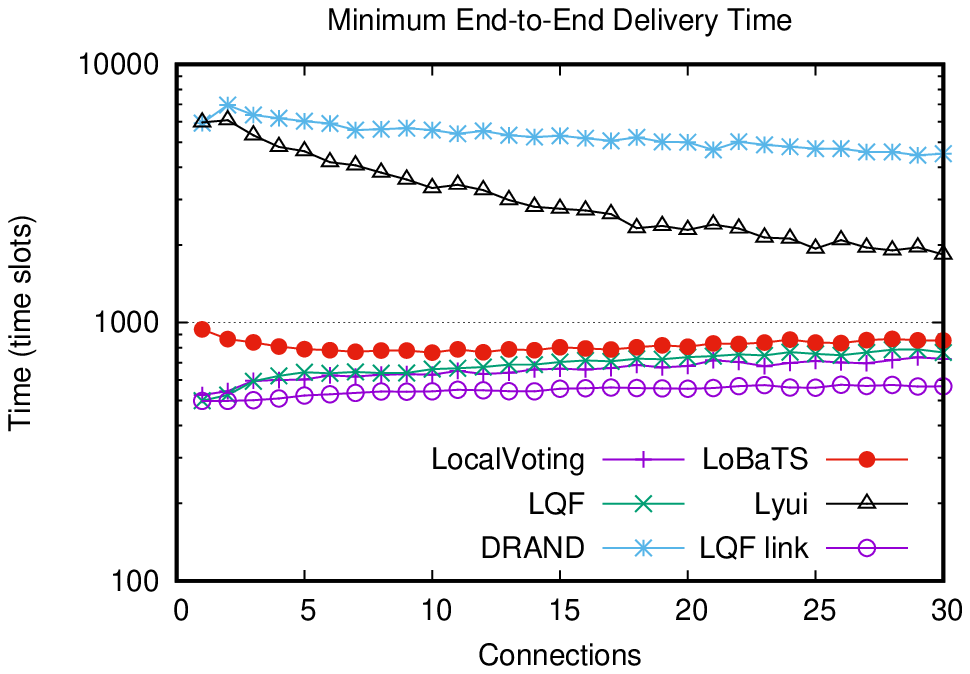}} 
	\caption{Simulation results of the different scheduling algorithms
		for different traffic load in the delivery time scenario.}
	\label{comp}
\end{figure*}

The results of the simulation are depicted in Fig.~\ref{comp}. For each number
of concurrent connections and each algorithm, the above metrics are averaged
over the $500$ different simulation runs.

Fig.~\ref{comp}(a) depicts the average end-to-end delivery times among all the
concurrent connections. The
\emph{LQF} and the \emph{Local Voting} algorithms achieve the shortest delivery
times, followed by \emph{LoBaTS}. The
\emph{DRAND} and \emph{Lyui} algorithms exhibit the worst performance, that is
expected, since these two algorithms assign a fixed number of slots to each
node without considering the traffic conditions. 
Fig.~\ref{comp}(b) presents the fairness in terms of the end-to-end delivery time
among connections that is calculated using Jain's fairness index. The \emph{LQF} and
\emph{Local Voting} algorithms clearly achieve superior fairness than other
algorithms, regardless of the number of concurrent connections. This illustrates the
significance of load balancing when considering fairness. The \emph{LoBaTS}
algorithm comes third (for most traffic loads) since it is also traffic
dependent, while the \emph{DRAND} follows it. \emph{Lyui's} algorithm has the 
worst fairness, and this validates
what is expected, since it assigns a different number of time slots according
to the nodes' color, without considering the traffic conditions. The lack of
fairness is noticeable for all algorithms except \emph{LQF} and \emph{Local
	Voting}, even when the number of connections is limited. As the number of
connections increases, fairness deteriorates for all algorithms, but the
difference in performance among the \emph{Local Voting} and \emph{LQF}
algorithms and the remaining algorithms increases as the traffic load increases. It should be
noted that even the \emph{LQF} algorithm cannot achieve perfect fairness, and
this is due to the different levels of congestion in various parts of the
network. Namely, flows that encounter no (or only limited) congestion on their path have shorter delivery times than flows that encounter congestion, and this
effect cannot be mitigated by scheduling policies alone.

Fig.~\ref{comp}(c)  demonstrates the maximum end-to-end delivery time, which is
the completion time of the connection that ends the latest. This is an important
metric because it shows after how much time the system has delivered all packets to their destination, thus, it is related to the capacity of the network.
The results confirm our expectations that the \emph{LQF} algorithms achieves the best performance. However, the performance of the \emph{Local Voting} algorithm is
very close to optimal. This validates the results of Section~\ref{network 
	model} that load balancing can decrease the overall delivery time. The slight 
difference among these two algorithms can be explained by two facts: 1) the 
\emph{Local Voting} algorithm is distributed, therefore, the delays in 
propagating the state affect its efficiency, and 2) slot exchange between two nodes is not always possible in real 
systems since allocations by 
other neighbors may cause a conflict, thus, it limits the amount of load 
balancing that is feasible. The \emph{LoBaTS} algorithm exhibits worse 
performance than the first two algorithms, possibly because it assigns at 
least one slot to each node, even if the node does not have traffic. \emph{DRAND} and 
\emph{Lyui's} algorithms perform equally badly, i.e. several orders of magnitude 
behind the rest of the algorithms. This is expected since 
both algorithms do not adapt the scheduling to traffic requirements.

Fig.~\ref{comp}(d) depicts the end-to-end delivery time for the connections with
the shortest delivery time. In general, the \emph{Local Voting} algorithm
has slightly
better performance in terms of the minimum delay compared to the other algorithms.


\subsection{The Effect of the Network Density}

\begin{figure*}[tpb]
	\centering
	\subfigure[The average delivery time ($10$ concurrent connections)]
	{\includegraphics[width=0.48\textwidth]{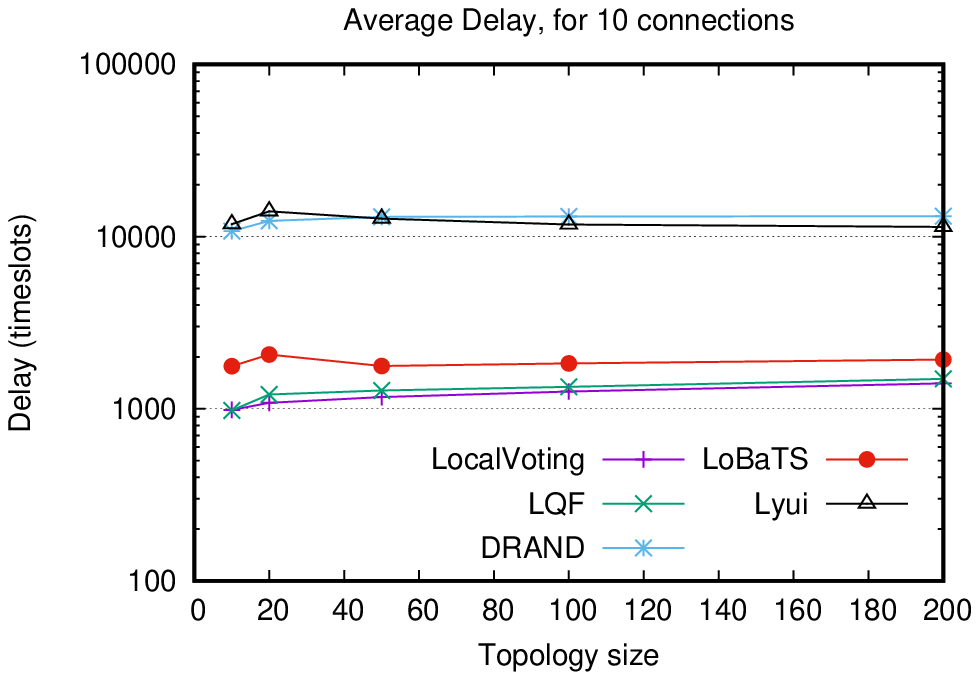}} 
	\hfill
	\subfigure[The average delivery time ($30$ concurrent connections)]
	{\includegraphics[width=0.48\textwidth]{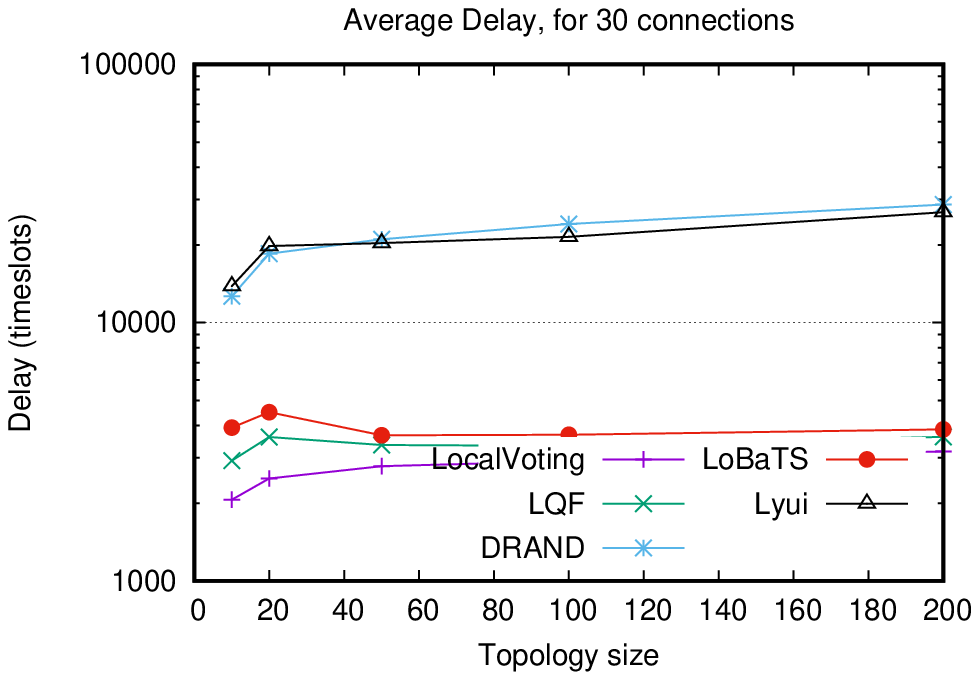}} 
	\caption{Simulation results for varying network density.}
	\label{fig:dens}
\end{figure*}

In this scenario we have repeated the experiments of section~\ref{sec:deliv}, but
this time we have changed the size of the topology, while the number of nodes is kept constant. This allows us to investigate how the network density affects the performance of the algorithms.

We vary the size of the network from $10$ units to $200$ units, while the number of nodes is still equal to $100$, and the transmission and the interference ranges are equal to
$10$ units. The results are depicted in Fig.~\ref{fig:dens} for $10$ and $30$ concurrent connections, respectively. In all cases the \emph{Local Voting} and \emph{LQF}
algorithms have the best performance. Additionally, the performance of the proposed \emph{Local Voting} algorithm is very
close to the performance of the centralized \emph{LQF} scheme in terms of maximum delivery
time.

\subsection{Steady State Scenario}

\begin{figure*}[tpb]
	\centering
	\subfigure[The average delivery time]
	{\includegraphics[width=0.48\textwidth]{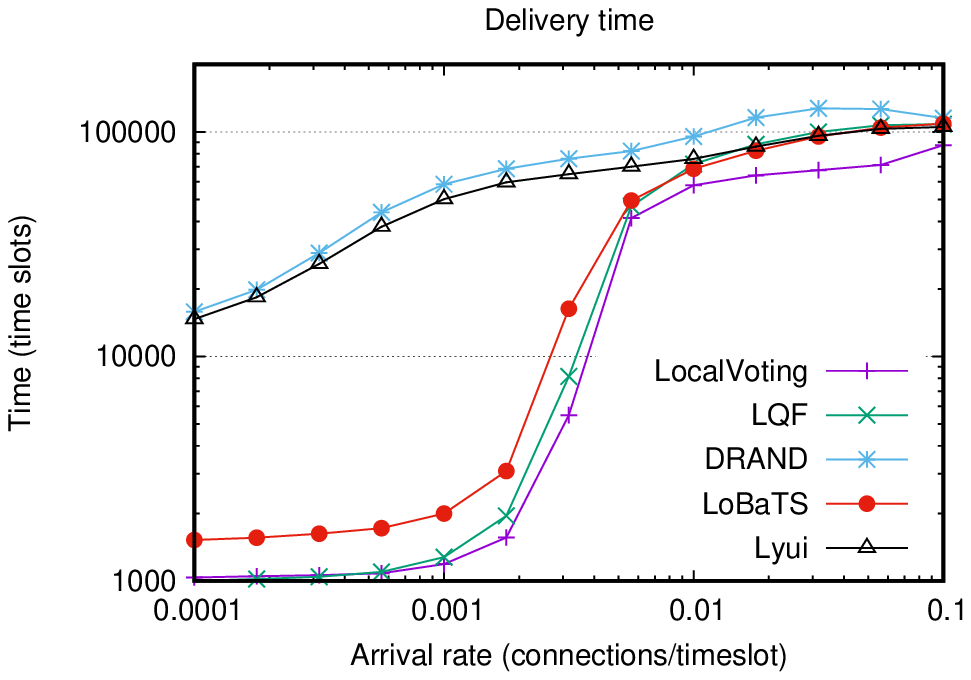}} 
	\hfill
	\subfigure[The average end-to-end packet delay]
	{\includegraphics[width=0.48\textwidth]{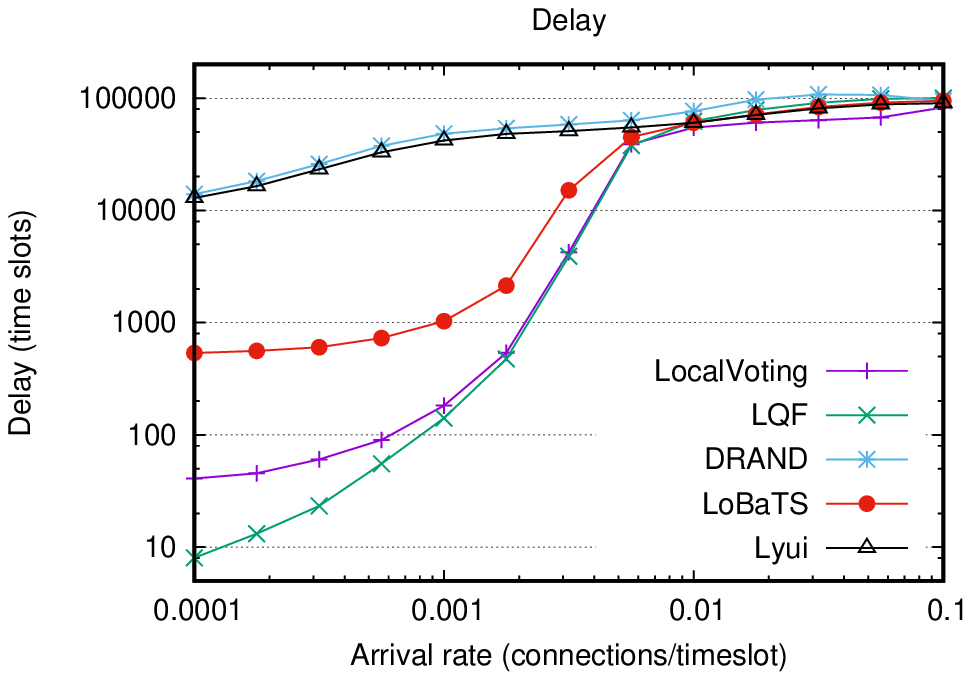}} 
	\subfigure[The average throughput]
	{\includegraphics[width=0.48\textwidth]{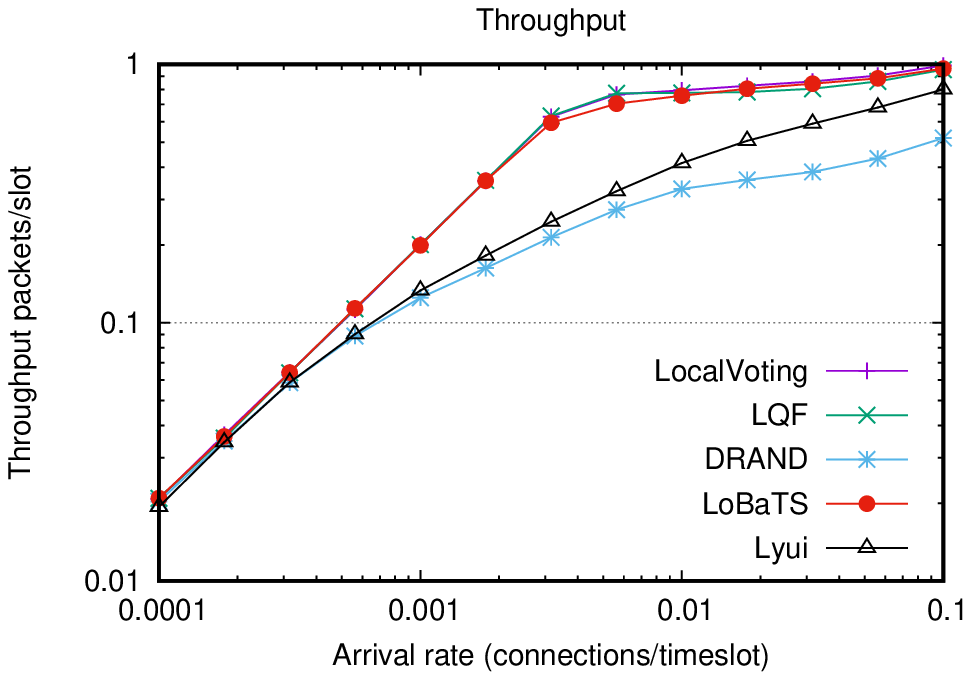}} 
	\hfill
	\subfigure[The fairness in terms of throughput]
	{\includegraphics[width=0.48\textwidth]{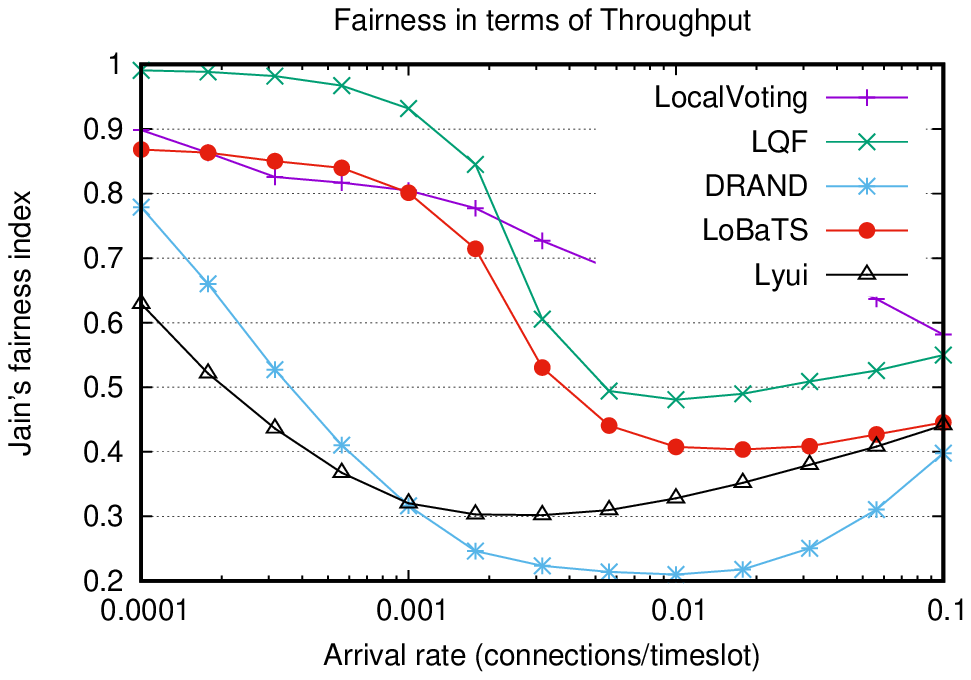}} 
	\caption{Simulation results for the steady state scenario.}
	\label{comp_steady}
\end{figure*}

In this subsection we evaluate the steady state performance of the load
balancing algorithm. This scenario is set up on the same network as the
previous one. However, instead of starting all connections at the beginning
of the simulation, the connections start following a Poisson process where the arrival rate $\lambda$ is in the range of
$\left[10^{-4}, 10^{-1}\right] \text{slots}^{-1}$, the duration of each
connection is distributed exponentially with a parameter of
$1/{\mu} = 10^{-3} \text{slots}^{-1}$, and the packet inter-arrival time within
a connection is $1$ packet every $5$ time slots. The source and the destination
of the connection are chosen randomly, following a uniform distribution. The
duration of the simulation is $3\times 10^6$ time slots. The packets that are
received before $36666$ slots have elapsed since the beginning of the simulation
are ignored.

\begin{figure*}[tpb]
	\centering
	\subfigure[Nodal delay]
	{\includegraphics[width=0.48\textwidth]{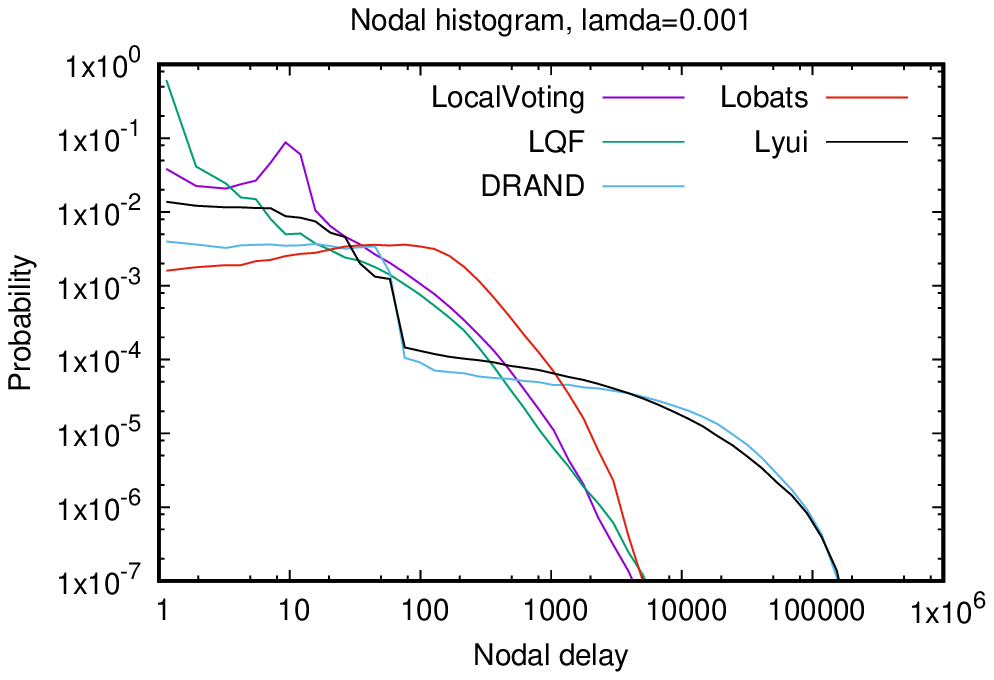}}
	\hfill
  \subfigure[End-to-end delay]{\includegraphics[width=0.48\textwidth]{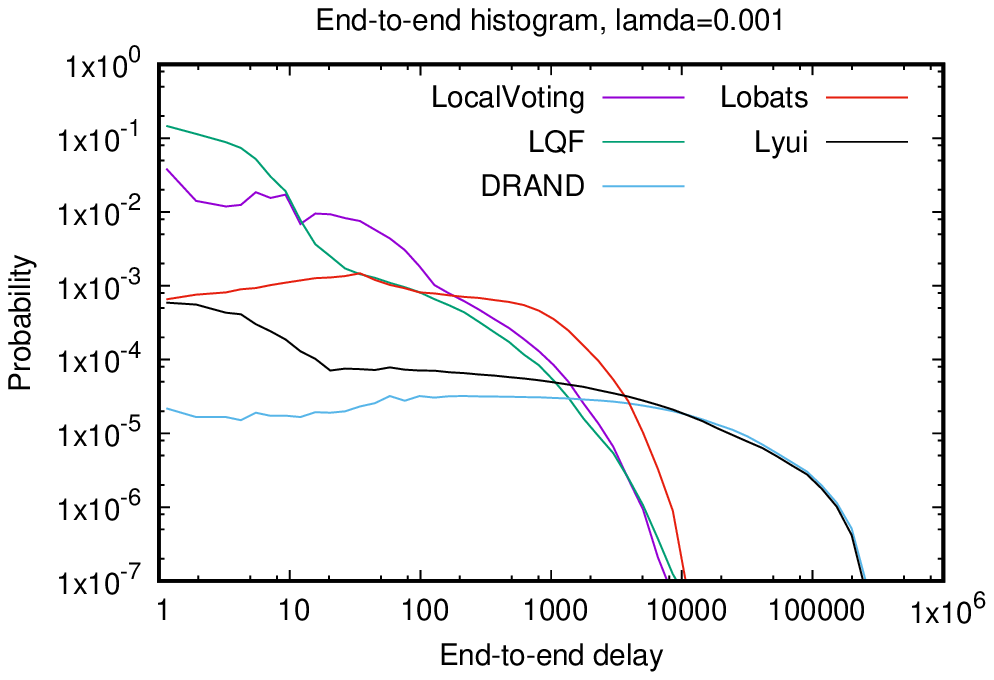}} 
	\caption{The distribution of delay per packet (nodal and end-to-end) for the
           steady state scenario.}
	\label{queue_steady}
\end{figure*}

We measure the average end-to-end delivery time, the average end-to-end delay,
the average throughput, and the fairness in terms of throughput.
Fig.~\ref{comp_steady}(a)  presents the average end-to-end delivery time, from
the transmission of the first packet to the reception of the last packet of all
connections. The \emph{Local Voting} algorithm achieves the best performance that is very close to the \emph{LQF} algorithm. The performance of the
\emph{LoBaTS} algorithm is a bit behind the first two algorithms, and
the traffic independent algorithms achieve the worst performance. In
Fig.~\ref{comp_steady}(b) we can see the average end-to-end delay, from the
moment a packet was generated until it was received by the final destination.
For low arrival rates, the \emph{LQF} algorithm has the smallest end-to-end
delay, followed by the \emph{Local Voting}, \emph{LoBaTS}, 
\emph{Lyui's} and \emph{DRAND} algorithms. On the contrary,
the average throughput for the \emph{LQF},  \emph{Local Voting}, and \emph{LoBaTS}
algorithms has a similar value, but \emph{Lyui's} and
\emph{DRAND} achieve lower average throughput (Fig.~\ref{comp_steady}(c)). Finally, in terms of
fairness, the \emph{Local Voting} algorithm is superior for medium arrival rates,
but \emph{LQF} has a superior performance for high and low arrival rates.

Fig.~\ref{queue_steady}(a) shows the evolution of the
delay per packet per node, for the different algorithms for an arrival rate of 
$10^{-3}$ new connections per time slot. 
The LQF algorithm has the higher percentage of packets with very low delay,
and this is expected because there is no frame length, so packets are eligible
to be transmitted at the next time slot. 
On the contrary, the \emph{Local Voting} algorithm has a peak in the delay
distribution that is close to the frame length of 10. 
The LoBaTS algorithm has higher delay, followed by DRAND and Lyui.

In Fig.~\ref{queue_steady}(b) we plot the distribution of the end-to-end delay
per packet. 
We can see that the ranking of the algorithms is similar to the per hop ranking.
This result validates that optimizing per-node delay through load balancing
has a positive effect on
end to end delay in a multihop network.

\subsection{The Effect of Packet Loss}

\begin{figure*}[tpb]
  \begin{minipage}[b]{0.48\textwidth}
	  \centering
    \includegraphics[width=\textwidth]{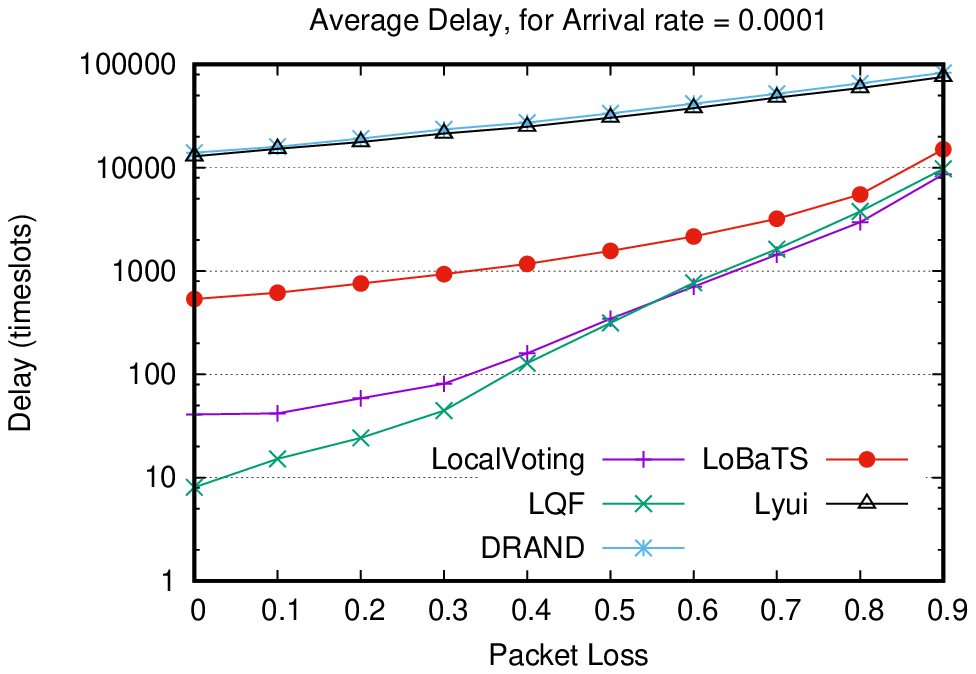}
  	\caption{Simulation results for different values of the packet loss, for an
	           arrival rate of $10^{-4}$ connections/slot.}
  	\label{loss4}
  \end{minipage}
  \hfill
  \begin{minipage}[b]{0.48\textwidth}
	  \centering
    \includegraphics[width=\textwidth]{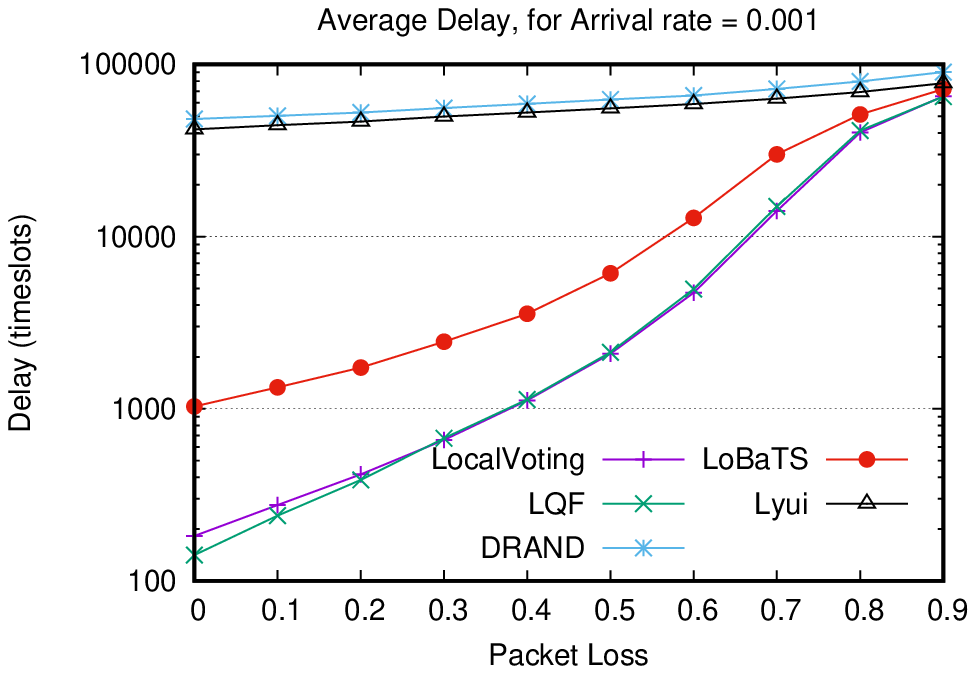} 
	  \caption{Simulation results for different values of the packet loss for an
		         arrival rate of $10^{-3}$ connections/slot.}
  	\label{loss3}
  \end{minipage}
\end{figure*}

\nop{
\begin{figure*}[thpb]
	\centering
	\subfigure[The average delivery time]
	{\includegraphics[width=0.48\textwidth]{expo_deliv04}} 
	\hfill
	\subfigure[The average end-to-end packet delay]
	{\includegraphics[width=0.48\textwidth]{expo_del04}} 
	\subfigure[The average throughput]
	{\includegraphics[width=0.48\textwidth]{expo_th04}} 
	\hfill
	\subfigure[The fairness in terms of throughput]
	{\includegraphics[width=0.48\textwidth]{expo_thf04}} 
	\caption{Simulation results for the steady state scenario. with packet loss
		         = $0.4$}
	\label{loss04}
\end{figure*}
}

\begin{figure*}[thpb]
	\centering
  \begin{minipage}[b]{0.48\textwidth}
    \includegraphics[width=\textwidth]{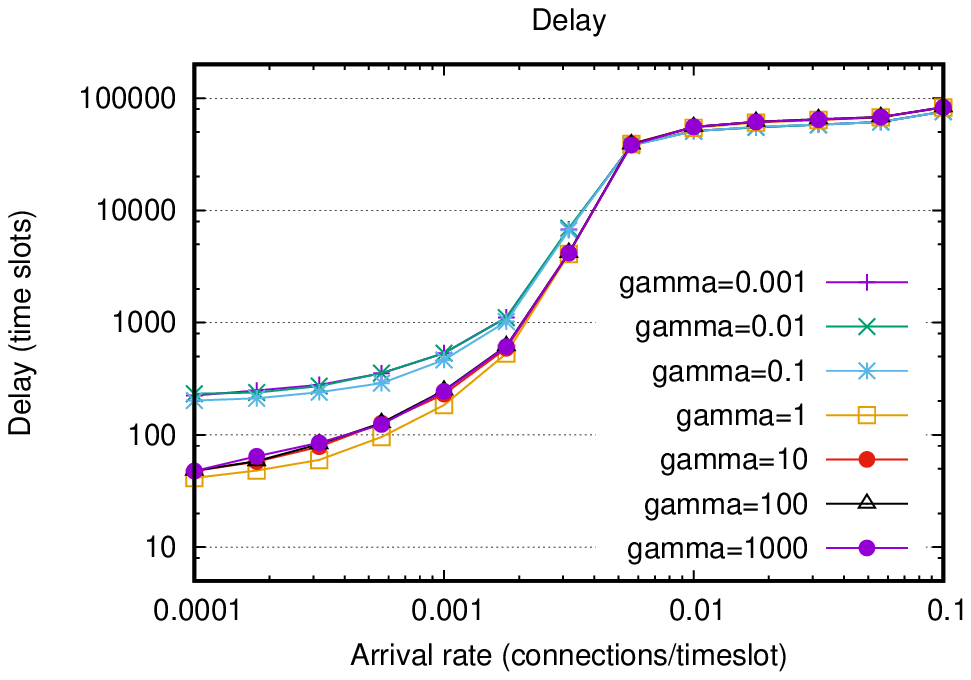} 
	  \caption{Simulation results for local voting protocol, on the steady state
  	   scenario, with $\gamma$ ranging from $10^{-3}$ to $10^3$.}
	  \label{gamma}
  \end{minipage}
  \hfill
  \begin{minipage}[b]{0.48\textwidth}
    \includegraphics[width=\textwidth]{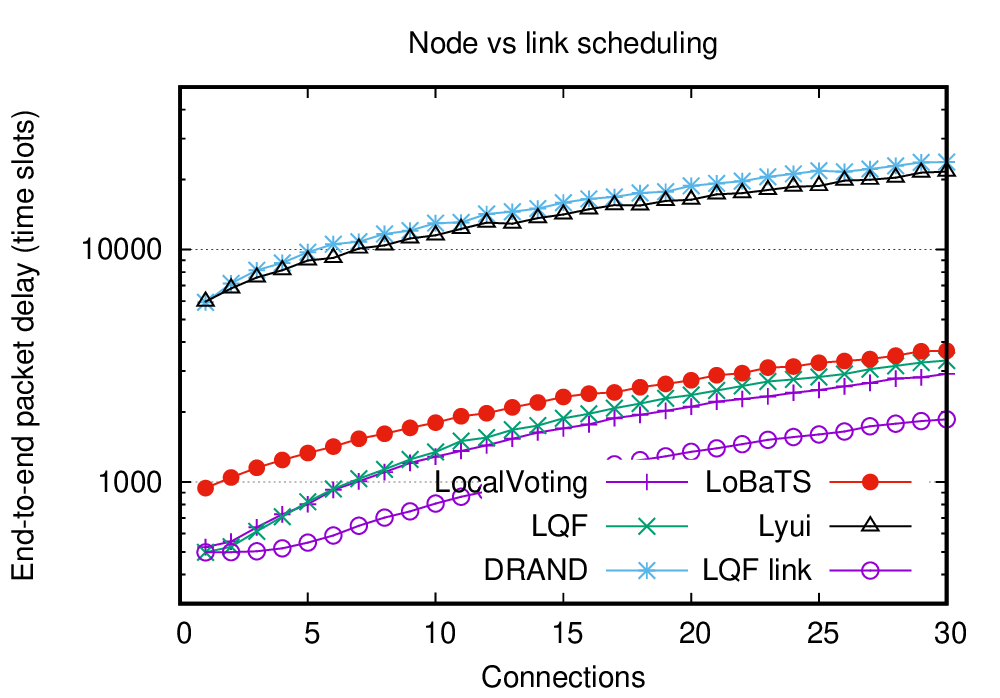} 
	  \caption{Node vs. Link scheduling}
	  \label{link}
  \end{minipage}
\end{figure*}

In this scenario, we evaluate the performance of the scheduling algorithms when errors can occur during the transmission between nodes. We kept the same
parameters as the previous scenario, but this time we considered a packet loss
probability in a range from zero (i.e. no packet loss) to $0.9$. We measure the
average delivery time, the average end-to-end delay, and the average throughput
for arrival rates of $10^{-4}$ and $10^{-3}$ connections per time slot.

Fig.~\ref{loss4} shows the results for an arrival rate equal to $10^{-4}$, and Fig.~\ref{loss3} presents the results for an arrival rate equal to $10^{-3}$
connections per time slot. In both cases, when the packet loss increases, the
end-to-end delay also increases. This is expected, because an increased packet loss
causes the packets to be re-transmitted, thus, an additional
delay is experienced. 
Similar results may be seen for the delivery time and the throughput, but are
omitted due to space page limitation.


\subsection{The Effect of the $\gamma$ Value}

In this scenario we investigate the effect of the $\gamma$ value on the
performance of the network. We execute the steady state scenario for
the \emph{Local Voting} algorithm, but this time, we set the $\gamma$ parameter
to different values, from $10^{-3}$ to $10^3$. The results are depicted in
Fig.~\ref{gamma}. 
There are significant differences in
terms of the end-to-end delay. 
For the network settings tested, we observed the best performance with in terms
of delay for $\gamma = 1$.


\subsection{Node Scheduling vs. Link Scheduling}
All the algorithms studied in this paper are node-scheduling algorithms.
This means that the destination of each transmission is not considered, so the 
interference model that is used under node-scheduling is more conservative
than link-scheduling. On the other hand, node scheduling has a multiplexing 
advantage under intermittent load. 
Fig.~\ref{link} depicts the results of the first scenario, including a 
link-scheduling variant of the LQF algorithm.

\section{Conclusion}
\label{conclusion}

The problem of scheduling is one of the big challenges in wireless networks. In
this paper we studied the interaction of scheduling and load balancing. We
showed that the problem of minimizing the overall delivery time through a
multihop network can be modeled as a consensus problem, where the goal is to
semi-equalize the fraction of the number of slots allocated to each node over the
queue length of the node. We introduced the schedule exchange graph, that is a
directed, time-varying graph, which represents whether a node can give a slot to
another node. The problem of wireless scheduling was modeled as a load balancing
problem. Taking into consideration the dynamically changing network topology, we
introduced \emph{Local Voting} protocol (consensus protocol) to solve the
scheduling/load balancing problem. Finally, we found the conditions that should be
met in order for the \emph{Local Voting} protocol to achieve approximate
consensus, and therefore optimize the delivery time throughout the network.

We compared the performance of the \emph{Local Voting} algorithm with other
scheduling algorithms from the literature. Simulation results validated the
theoretical analysis and showed that the delivery times are minimized with the use of the \emph{Local Voting} algorithm. The proposed algorithm achieves
better performance than the other known distributed algorithms from the
literature in terms of the average delay, the  maximum delay, and the fairness. Despite
being distributed, the performance of the \emph{Local Voting} algorithm is very close to the performance of the centralized \emph{LQF} algorithm which is considered to
have the best performance. To summarize, we showed the advantage of load balancing
when performing scheduling in wireless multihop networks, proposed \emph{Local
	Voting} algorithm for load balancing/scheduling, found theoretical conditions
for convergence (reaching consensus), and demonstrated by simulations that
the \emph{Local Voting} algorithm shows good performance in comparison with other
scheduling algorithms.

\section*{Appendix}
\label{6-appendix}

\section{Proof of Theorem 1}

\begin{IEEEproof}
The result of this Theorem and its proof are different from corresponding parts in~\cite{7047923}. The difference is caused by the different ways of achieving consensus. While in~\cite{7047923}, consensus is achieved through re-distributing the load or $q_t^i$, in this paper consensus is reached through re-distributing slots in a frame, i.e. $p_t^i$.
The idea of the proof follows the paper~\cite{Amelina13CDC}.

By virtue the Eqs.~(\ref{Nat_11}) and (\ref{eq_w}), the dynamics $p_t^i$ of  the closed loop system with protocol~(\ref{Nat_7}) are as
follows
$$
p^i_{t+1} = p^i_{t} + e^i_{t+1}+ [ \gamma \sum_{\tilde N_{t}^{i}} a_t^{i, j}  (\tilde x^j_t - \tilde x^i_t)] =
$$
\begin{equation}
\label{Nat_12}
 p^i_{t} +  \gamma \sum_{\tilde N_{t}^{i}} a_t^{i, j}  (\tilde x^j_t - \tilde x^i_t) + e^i_{t+1}+ w_t^i.
\end{equation}

Denote by ${{\mathbf{p}}}_{t} \in {\mathbb R}^{n}$ a vector which consists of $p_t^i$,  ${{\mathbf{e}}}_{t+1} \in {\mathbb R}^{n}$ a vector which consists of $e_{t+1}^i$, and by ${{\mathbf{w}}}_{t} \in {\mathbb R}^{n}$ a vector of the errors $w_t^i$, where $t=0, 1, \ldots$.

Due to the view of the Laplacian matrix ${\cal L}({A}_t)$ and definition of $Q_{t+1}$, we can rewrite Eq.~(\ref{Nat_12}) in a vector-matrix form as:
\begin{equation}
\label{Nat_8}
{ {\mathbf{p}}}_{t+1} = { {\mathbf{p}}}_{t} - \gamma {\cal L}({A}_t)Q_{t+1}^{-1}{ {\mathbf{p}}}_t + { {\mathbf{e}}}_{t+1}+{ {\mathbf{w}}}_{t}.
\end{equation}

We consider that $\bar {\mathbf{p}}_{t} =  Q_{av} \bar {\mathbf{x}}_{t}$. If we
multiply both sides of Eq. (\ref{aver_syst}) by $Q_{av} $, we get that the
sequence $\{{\bar {\mathbf{p}}}_{t} \}$ is a trajectory of the average system
\begin{equation}\label{aver_syst_p}
    {\bar {\mathbf{p}}}_{t+1} = {\bar {\mathbf{p}}}_{t} - \gamma {\cal L}({B}_{av}){\bar {\mathbf{p}}}_t + {\bar {\mathbf{e}}}_{t+1}.
\end{equation}

The vector $\mathbf{1}_{ n}$ is the right eigenvector of the Laplacian-type matrices
${\tilde{\cal L}}_t= \gamma{\cal L}({ A}_t)Q_{t+1}^{-1}=\gamma{\cal L}({ B}_t) $ and ${\bar{\cal L}}_B = \gamma{\cal L}(B_{av}) $ corresponding to the zero eigenvalue: ${\tilde{\cal L}}_t \mathbf{1}_{ n} \mathbf{1}_{ n} = {\bar{\cal L}}_B \mathbf{1}_{ n} = 0$.
Sums of all elements in the rows of the matrices  ${\tilde{\cal L}}_t$ or ${\bar{\cal L}}_B$
are equal to zero and, moreover, all the diagonal elements are positive and equal
to the absolute value of the sum of all other elements in the row.

The next Lemma from~\cite{ren2005consensus} is useful.

{\bf Lemma} \cite{ren2005consensus}: Laplacian matrix ${\cal L}(B)$ of graph ${\cal G}_B$ has an algebraic multiplicity equal to
one for its eigenvalue $\lambda_1=0$ if and only if   graph ${\cal G}_B$ has a
spanning tree.

Note that graph $ {\cal G}_{B_{av}}$ has a spanning tree when conditions  {\bf A1.c} and {\bf A2} hold.


Due to the definitions
of the matrices ${\bar{\cal L}}_t$ and ${\tilde{\cal L}}_A$,
we derive from (\ref{Nat_8}),(\ref{aver_syst_p})  for the difference
${{\mathbf{r}}}_{t+1}= {{\mathbf{p}}}_{t+1} - {\bar{\mathbf{p}}}_{t+1}$
$$
 {{\mathbf{r}}}_{t+1} =  {{\mathbf{r}}}_{t} - {\tilde{\cal L}}_t{{\mathbf{p}}}_{t}  + {\bar{\cal L}}_B{\bar {\mathbf{p}}}_{t} + {\mathbf{e}}_{t+1} - {\bar{\mathbf{e}}}_{t+1} + {\mathbf{w}}_{t}=
 $$
 $$
 =  ({ I}- {\bar{\cal L}}_B){{\mathbf{r}}}_{t} -
 ({\bar{\cal L}}_t-{\tilde{\cal L}}_B ){{{\mathbf{p}}}}_t + ({\mathbf{e}}_{t+1}- \bar {\mathbf{e}}_{t+1})+ {\mathbf{w}}_{t},
$$
where ${I}$ is the identity matrix.

 Consider the conditional mathematical expectation of the squared norm ${{\mathbf{r}}}_{t+1}$ according to $\sigma$-algebra ${{\cal F}_t}$.
 By virtue of Assumptions {\bf A1.d--f} we derive
$$
    {\mathrm{E}}_{{\cal F}_t} \|{{\mathbf{r}}}_{t+1}\|^2 \leq \|({ I}-{\tilde{\cal L}}_B) {{\mathbf{r}}}_{t} \|^2 +
    {{{\mathbf{p}}}}_t^{\rm T}R{{{\mathbf{p}}}}_t + n\sigma_e^2 + n\sigma_w^2.$$

Further, by  taking
 unconditional expectation  we get:
$
    {\mathrm{E}} \|{{\mathbf{r}}}_{t+1}\|^2 \leq \rho {\mathrm{E}} \|{{\mathbf{r}}}_{t} \|^2 +
    \Delta.
$
By Lemma~1 from Chapter~2 of~\cite{PolyakBook} it follows that
\begin{equation}\label{p_bound}
 {\mathrm{E}} \|{{\mathbf{r}}}_{t+1}\|^2  \leq  \frac{\Delta}{1-\rho}+  \rho^t {\mathrm{E}} \|{{\mathbf{p}}}_{0} \|^2.
\end{equation}

Due to the definitions we have
$$
{\rm E}\|{{\mathbf{x}}}_{t+1} - {\bar{\mathbf{x}}}_{t+1}\|^2 = {\rm E}\| Q_{t+1}^{-1}( {{\mathbf{p}}}_{t+1} - {\bar{\mathbf{p}}}_{t+1}) + (Q_{t+1}^{-1} Q_{av} - I){\bar{\mathbf{x}}}_{t+1}\|^2 \leq
$$
$$
2{\rm E} \| Q_{t+1}^{-1}{{\mathbf{r}}}_{t+1}\|^2 +2{\rm E}\| (Q_{t+1}^{-1}  - Q_{av}^{-1})Q_{av}{\bar{\mathbf{x}}}_{t+1}\|^2 \leq
$$
$$
2 (\frac{\Delta}{1-\rho}+  \rho^t {\mathrm{E}} \|{{\mathbf{p}}}_{0} \|^2) +2\sigma_q^2\| Q_{av} {\bar{\mathbf{x}}}_{t+1}\|^2.
$$
The proof of the
 first part of Theorem 1 is completed.

The second conclusion about the asymptotic mean square $\varepsilon$-consensus follows from inequality~(\ref{Nat_T1}) if $t \to \infty$. Since (\ref{cond_gamma2})  is satisfied, then the third term of~(\ref{Nat_T1}) exponentially tends to zero.

\end{IEEEproof}

\section*{Acknowledgment}

This work was supported 
by RFBR under Grants No.15-08-02640 and No.16-07-00890. 
We would like to thank the anonymous reviewers for their 
very valuable comments.

\bibliography{SchedulingConsensus}

\begin{thebibliography}{10}
\providecommand{\url}[1]{#1}
\csname url@samestyle\endcsname
\providecommand{\newblock}{\relax}
\providecommand{\bibinfo}[2]{#2}
\providecommand{\BIBentrySTDinterwordspacing}{\spaceskip=0pt\relax}
\providecommand{\BIBentryALTinterwordstretchfactor}{4}
\providecommand{\BIBentryALTinterwordspacing}{\spaceskip=\fontdimen2\font plus
\BIBentryALTinterwordstretchfactor\fontdimen3\font minus
  \fontdimen4\font\relax}
\providecommand{\BIBforeignlanguage}[2]{{%
\expandafter\ifx\csname l@#1\endcsname\relax
\typeout{** WARNING: IEEEtran.bst: No hyphenation pattern has been}%
\typeout{** loaded for the language `#1'. Using the pattern for}%
\typeout{** the default language instead.}%
\else
\language=\csname l@#1\endcsname
\fi
#2}}
\providecommand{\BIBdecl}{\relax}
\BIBdecl

\bibitem{kiess2007survey}
W.~Kiess and M.~Mauve, ``A survey on real-world implementations of mobile
  ad-hoc networks,'' \emph{Ad Hoc Networks}, vol.~5, no.~3, pp. 324--339, 2007.

\bibitem{pottie1998wireless}
G.~J. Pottie, ``Wireless sensor networks,'' in \emph{Information Theory
  Workshop, 1998}.\hskip 1em plus 0.5em minus 0.4em\relax IEEE, 1998, pp.
  139--140.

\bibitem{akyildiz2005wireless}
I.~F. Akyildiz, X.~Wang, and W.~Wang, ``Wireless mesh networks: a survey,''
  \emph{Computer Networks}, vol.~47, no.~4, pp. 445--487, 2005.

\bibitem{bettstetter2005connectivity}
C.~Bettstetter and C.~Hartmann, ``Connectivity of wireless multihop networks in
  a shadow fading environment,'' \emph{Wireless Networks}, vol.~11, no.~5, pp.
  571--579, 2005.

\bibitem{gupta1998critical}
P.~Gupta and P.~R. Kumar, ``Critical power for asymptotic connectivity in
  wireless networks,'' in \emph{Stochastic Analysis, Control, Optimization and
  Applications}.\hskip 1em plus 0.5em minus 0.4em\relax Springer, 1998, pp.
  547--566.

\bibitem{gupta2000capacity}
------, ``The capacity of wireless networks,'' \emph{IEEE Trans. Inf. Theor.},
  vol.~46, no.~2, pp. 388--404, Sep. 2006.

\bibitem{li2001capacity}
J.~Li, C.~Blake, D.~S. De~Couto, H.~I. Lee, and R.~Morris, ``Capacity of ad hoc
  wireless networks,'' in \emph{Proceedings of the 7th Annual International
  Conference on Mobile Computing and Networking}.\hskip 1em plus 0.5em minus
  0.4em\relax ACM, 2001, pp. 61--69.

\bibitem{grossglauser2002mobility}
M.~Grossglauser and D.~N. Tse, ``Mobility increases the capacity of ad hoc
  wireless networks,'' \emph{Networking, IEEE/Acm Transactions on}, vol.~10,
  no.~4, pp. 477--486, 2002.

\bibitem{weber2010overview}
S.~Weber, J.~G. Andrews, and N.~Jindal, ``An overview of the transmission
  capacity of wireless networks,'' \emph{Communications, IEEE Transactions on},
  vol.~58, no.~12, pp. 3593--3604, 2010.

\bibitem{tassiulas1992stability}
L.~Tassiulas and A.~Ephremides, ``Stability properties of constrained queueing
  systems and scheduling policies for maximum throughput in multihop radio
  networks,'' \emph{Automatic Control, IEEE Transactions on}, vol.~37, no.~12,
  pp. 1936--1948, 1992.

\bibitem{lin2004joint}
X.~Lin and N.~B. Shroff, ``Joint rate control and scheduling in multihop
  wireless networks,'' in \emph{Decision and Control, 2004. CDC. 43rd IEEE
  Conference on}, vol.~2.\hskip 1em plus 0.5em minus 0.4em\relax IEEE, 2004,
  pp. 1484--1489.

\bibitem{salem2005fair}
N.~B. Salem and J.-P. Hubaux, ``A fair scheduling for wireless mesh networks,''
  in \emph{Proc. IEEE Workshop on Wireless Mesh Networks (WiMesh)}, 2005.

\bibitem{Ning2012533}
Z.~Ning, L.~Guo, Y.~Peng, and X.~Wang, ``Joint scheduling and routing algorithm
  with load balancing in wireless mesh network,'' \emph{Computers \& Electrical
  Engineering}, vol.~38, no.~3, pp. 533--550, 2012.

\bibitem{li2007joint}
Y.~Li and A.~Ephremides, ``A joint scheduling, power control, and routing
  algorithm for ad hoc wireless networks,'' \emph{Ad Hoc Networks}, vol.~5,
  no.~7, pp. 959--973, 2007.

\bibitem{cruz2003optimal}
R.~L. Cruz and A.~V. Santhanam, ``Optimal routing, link scheduling and power
  control in multihop wireless networks,'' in \emph{INFOCOM 2003. Twenty-Second
  Annual Joint Conference of the IEEE Computer and Communications. IEEE
  Societies}, vol.~1.\hskip 1em plus 0.5em minus 0.4em\relax IEEE, 2003, pp.
  702--711.

\bibitem{hyytia2006load}
E.~Hyytia and J.~Virtamo, ``On load balancing in a dense wireless multihop
  network,'' in \emph{Next Generation Internet Design and Engineering, 2006.
  NGI'06. 2006 2nd Conference on}.\hskip 1em plus 0.5em minus 0.4em\relax IEEE,
  2006, pp. 8--pp.

\bibitem{sgora2015survey}
A.~Sgora, D.~J. Vergados, and D.~D. Vergados, ``A survey of tdma scheduling
  schemes in wireless multihop networks,'' \emph{ACM Computing Surveys (CSUR)},
  vol.~47, no.~3, p.~53, 2015.

\bibitem{gunasekaran2010efficient}
R.~Gunasekaran, S.~Siddharth, P.~Krishnaraj, M.~Kalaiarasan, and V.~R.
  Uthariaraj, ``Efficient algorithms to solve broadcast scheduling problem in
  wimax mesh networks,'' \emph{Computer Communications}, vol.~33, no.~11, pp.
  1325--1333, 2010.

\bibitem{li2012efficient}
J.-S. Li, K.-H. Liu, and C.-H. Wu, ``Efficient group multicast node scheduling
  schemes in multi-hop wireless networks,'' \emph{Computer Communications},
  vol.~35, no.~10, pp. 1247--1258, 2012.

\bibitem{chiang2014decentralized}
C.-T. Chiang, H.-C. Chen, W.-H. Liao, and K.-P. Shih, ``A decentralized
  minislot scheduling protocol (dmsp) in tdma-based wireless mesh networks,''
  \emph{Journal of Network and Computer Applications}, vol.~37, pp. 206--215,
  2014.

\bibitem{arivudainambi2014heuristic}
D.~Arivudainambi and D.~Rekha, ``Heuristic approach for broadcast scheduling,
  problem in wireless mesh networks,'' \emph{AEU-International Journal of
  Electronics and Communications}, vol.~68, no.~6, pp. 489--495, 2014.

\bibitem{liu2012topology}
Y.~Liu, V.~O. Li, K.-C. Leung, and L.~Zhang, ``Topology-transparent distributed
  multicast and broadcast scheduling in mobile ad hoc networks,'' in
  \emph{Vehicular Technology Conference (VTC Spring), 2012 IEEE 75th}.\hskip
  1em plus 0.5em minus 0.4em\relax IEEE, 2012, pp. 1--5.

\bibitem{zeng2014collaboration}
B.~Zeng and Y.~Dong, ``A collaboration-based distributed tdma scheduling
  algorithm for data collection in wireless sensor networks,'' \emph{Journal of
  Networks}, vol.~9, no.~9, pp. 2319--2327, 2014.

\bibitem{xu2011topology}
C.~Xu, Y.~Xu, Z.~Wang, and H.~Luo, ``A topology-transparent mac scheduling
  algorithm with guaranteed qos for multihop wireless network,'' \emph{Journal
  of Control Theory and Applications}, vol.~9, no.~1, pp. 106--114, 2011.

\bibitem{lam2014broadcast}
N.~Lam, M.~K. An, D.~T. Huynh, and T.~Nguyen, ``Broadcast scheduling problem in
  sinr model,'' \emph{International Journal of Foundations of Computer
  Science}, vol.~25, no.~03, pp. 331--342, 2014.

\bibitem{Arivudainambi2016}
D.~Arivudainambi and S.~Balaji, ``Improved memetic algorithm for energy
  efficient sensor scheduling with adjustable sensing range,'' \emph{Wireless
  Personal Communications}, pp. 1--22, 2016.

\bibitem{6777543}
Y.~Liu, V.~O.~K. Li, K.~C. Leung, and L.~Zhang, ``Performance improvement of
  topology-transparent broadcast scheduling in mobile ad hoc networks,''
  \emph{IEEE Transactions on Vehicular Technology}, vol.~63, no.~9, pp.
  4594--4605, Nov 2014.

\bibitem{7524609}
X.~Tian, J.~Yu, L.~Ma, G.~Li, and X.~Cheng, ``Distributed deterministic
  broadcasting algorithms under the sinr model,'' in \emph{IEEE INFOCOM}, April
  2016, pp. 1--9.

\bibitem{6824752}
J.~G. Andrews, S.~Buzzi, W.~Choi, S.~V. Hanly, A.~Lozano, A.~C.~K. Soong, and
  J.~C. Zhang, ``What will 5g be?'' \emph{IEEE Journal on Selected Areas in
  Communications}, vol.~32, no.~6, pp. 1065--1082, June 2014.

\bibitem{Panwar201664}
\BIBentryALTinterwordspacing
N.~Panwar, S.~Sharma, and A.~K. Singh, ``A survey on 5g: The next generation of
  mobile communication,'' \emph{Physical Communication}, vol. 18, Part 2, pp.
  64 -- 84, 2016, special Issue on Radio Access Network Architectures and
  Resource Management for 5G. [Online]. Available:
  \url{http://www.sciencedirect.com/science/article/pii/S1874490715000531}
\BIBentrySTDinterwordspacing

\bibitem{Li:2013:OMB:2509723}
J.~Li, X.~Wu, and R.~Laroia, \emph{OFDMA Mobile Broadband Communications: A
  Systems Approach}, 1st~ed.\hskip 1em plus 0.5em minus 0.4em\relax New York,
  NY, USA: Cambridge University Press, 2013.

\bibitem{7752514}
J.~Xiao, C.~Yang, J.~Wang, and H.~Dai, ``Joint interference management in
  ultra-dense small cell networks: A multi-dimensional coordination,'' in
  \emph{2016 8th International Conference on Wireless Communications Signal
  Processing (WCSP)}, Oct 2016, pp. 1--5.

\bibitem{7600897}
M.~A. Gutierrez-Estevez, D.~Gozalvez-Serrano, M.~Botsov, and S.~Staczak,
  ``Stfdma: A novel technique for ad-hoc v2v networks exploiting radio channels
  frequency diversity,'' in \emph{2016 International Symposium on Wireless
  Communication Systems (ISWCS)}, Sept 2016, pp. 182--187.

\bibitem{6812287}
J.~G. Andrews, S.~Singh, Q.~Ye, X.~Lin, and H.~S. Dhillon, ``An overview of
  load balancing in hetnets: old myths and open problems,'' \emph{IEEE Wireless
  Communications}, vol.~21, no.~2, pp. 18--25, April 2014.

\bibitem{Niu2015}
Y.~Niu, Y.~Li, D.~Jin, L.~Su, and A.~V. Vasilakos, ``A survey of millimeter
  wave communications (mmwave) for 5g: opportunities and challenges,''
  \emph{Wireless Networks}, vol.~21, no.~8, pp. 2657--2676, 2015.

\bibitem{vergados2017local}
D.~J. Vergados, N.~Amelina, Y.~Jiang, K.~Kralevska, and O.~Granichin, ``Local
  voting: Optimal distributed node scheduling algorithm for multihop wireless
  networks,'' in \emph{INFOCOM Workshop Proceedings, Atlanta, GA, USA, 1-4 May
  2017}, 2017, pp. 931--932.

\bibitem{7047923}
N.~Amelina, A.~Fradkov, Y.~Jiang, and D.~J. Vergados, ``Approximate consensus
  in stochastic networks with application to load balancing,'' \emph{IEEE
  Transactions on Information Theory}, vol.~61, no.~4, pp. 1739--1752, April
  2015.

\bibitem{Tsitsiklis}
J.~Tsitsiklis, D.~Bertsekas, and M.~Athans, ``Distributed asynchronous
  deterministic and stochastic gradient optimization algorithms,''
  \emph{Automatic Control, IEEE Transactions on}, vol.~31, no.~9, pp. 803--812,
  1986.

\bibitem{Huang}
M.~Huang, ``Stochastic approximation for consensus: a new approach via ergodic
  backward products,'' \emph{IEEE Transactions on Automatic Control}, vol.~57,
  no.~12, pp. 2994--3008, 2012.

\bibitem{Borkar08}
V.~Borkar, \emph{Stochastic Approximation: a Dynamical Systems
  Viewpoint}.\hskip 1em plus 0.5em minus 0.4em\relax Cambridge University Press
  Cambridge, 2008.

\bibitem{GranAmelinaTAC15}
O.~Granichin and N.~Amelina, ``Simultaneous perturbation stochastic
  approximation for tracking under unknown but bounded disturbances,''
  \emph{IEEE Transactions on Automatic Control}, vol.~60, no.~6, pp.
  1653--1658, 2015.

\bibitem{chvatal1984perfectly}
V.~Chv{\'a}tal, ``Perfectly ordered graphs,'' \emph{North-Holland mathematics
  studies}, vol.~88, pp. 63--65, 1984.

\bibitem{Jain:2003:IIM:938985.938993}
\BIBentryALTinterwordspacing
K.~Jain, J.~Padhye, V.~N. Padmanabhan, and L.~Qiu, ``Impact of interference on
  multi-hop wireless network performance,'' in \emph{Proceedings of the 9th
  Annual International Conference on Mobile Computing and Networking}, ser.
  MobiCom '03.\hskip 1em plus 0.5em minus 0.4em\relax New York, NY, USA: ACM,
  2003, pp. 66--80. [Online]. Available:
  \url{http://doi.acm.org/10.1145/938985.938993}
\BIBentrySTDinterwordspacing

\bibitem{kashyap2007quantized}
A.~Kashyap, T.~Ba{\c{s}}ar, and R.~Srikant, ``Quantized consensus,''
  \emph{Automatica}, vol.~43, no.~7, pp. 1192--1203, 2007.

\bibitem{kar2010distributed}
S.~Kar and J.~M. Moura, ``Distributed consensus algorithms in sensor networks:
  Quantized data and random link failures,'' \emph{Signal Processing, IEEE
  Transactions on}, vol.~58, no.~3, pp. 1383--1400, 2010.

\bibitem{rhee2006drand}
I.~Rhee, A.~Warrier, J.~Min, and L.~Xu, ``Drand: distributed randomized tdma
  scheduling for wireless ad-hoc networks,'' in \emph{Proceedings of the 7th
  ACM international symposium on Mobile ad hoc networking and computing}.\hskip
  1em plus 0.5em minus 0.4em\relax ACM, 2006, pp. 190--201.

\bibitem{vergados2012fair}
D.~J. Vergados, A.~Sgora, D.~D. Vergados, D.~Vouyioukas, and
  I.~Anagnostopoulos, ``Fair tdma scheduling in wireless multihop networks,''
  \emph{Telecommunication Systems}, vol.~50, no.~3, pp. 181--198, 2012.

\bibitem{Jain:1984:QMF}
R.~Jain, D.~Chiu, and W.~Hawe, ``A quantitative measure of fairness and
  discrimination for resource allocation in shared computer systems,'' Digital
  Equipment Corporation, Maynard, MA, USA, DEC Research Report TR-301, Sep.
  1984.

\bibitem{Lyui}
W.-P. Lyui, ``Design of a new operational structure for mobile radio
  networks,'' \emph{Ph.D. dissertation, Clemson Univ., Clemson, SC}, 1991.

\bibitem{hammond2004properties}
J.~L. Hammond and H.~B. Russell, ``Properties of a transmission assignment
  algorithm for multiple-hop packet radio networks,'' \emph{Wireless
  Communications, IEEE Transactions on}, vol.~3, no.~4, pp. 1048--1052, 2004.

\bibitem{wolf2006distributed}
B.~J. Wolf, J.~L. Hammond, and H.~B. Russell, ``A distributed load-based
  transmission scheduling protocol for wireless ad hoc networks,'' in
  \emph{Proceedings of the 2006 International Conference on Wireless
  Communications and Mobile Computing}.\hskip 1em plus 0.5em minus 0.4em\relax
  ACM, 2006, pp. 437--442.

\bibitem{dimakis2006sufficient}
A.~Dimakis and J.~Walrand, ``Sufficient conditions for stability of
  longest-queue-first scheduling: Second-order properties using fluid limits,''
  \emph{Advances in Applied Probability}, pp. 505--521, 2006.

\bibitem{joo2009understanding}
C.~Joo, X.~Lin, and N.~B. Shroff, ``Understanding the capacity region of the
  greedy maximal scheduling algorithm in multihop wireless networks,''
  \emph{IEEE/ACM Transactions on Networking (TON)}, vol.~17, no.~4, pp.
  1132--1145, 2009.

\bibitem{Amelina13CDC}
N.~Amelina, O.~Granichin, and A.~Kornivetc, ``Local voting protocol in
  decentralized load balancing problem with switched topology, noise, and
  delays,'' \emph{Proc. of 52nd IEEE Conference on Decision and Control (CDC
  2013)}, pp. 4613--4618, 2013.

\bibitem{ren2005consensus}
W.~Ren and R.~W. Beard, ``Consensus seeking in multiagent systems under
  dynamically changing interaction topologies,'' \emph{Automatic Control, IEEE
  Transactions on}, vol.~50, no.~5, pp. 655--661, 2005.

\bibitem{PolyakBook}
B.~T. Polyak, \emph{Introduction to Optimization}.\hskip 1em plus 0.5em minus
  0.4em\relax Optimization Software, 1987.

\end{thebibliography}
\bibliographystyle{IEEEtran}

\ifCLASSOPTIONcaptionsoff
  \newpage
\fi



%

\end{document}